\pdfoutput=1
\documentclass[lettersize,journal]{IEEEtran}
\usepackage{amsmath,amsfonts}
\usepackage{algorithmic}
\usepackage{array}
\usepackage[caption=false,font=normalsize,labelfont=sf,textfont=sf]{subfig}
\usepackage{textcomp}
\usepackage{stfloats}
\usepackage{url}
\usepackage{verbatim}
\usepackage{graphicx}
\usepackage{pifont}
\usepackage{calc}
\usepackage{adjustbox}
\usepackage{ragged2e}
\usepackage{tabularx}

\def\BibTeX{{\rm B\kern-.05em{\sc i\kern-.025em b}\kern-.08em
    T\kern-.1667em\lower.7ex\hbox{E}\kern-.125emX}}
\usepackage{balance}
\begin{document}
\title{Noise Suppression and Radio Frequency Interference Rejection for 
Self-Triggered Radio Detectors of Extensive Air Showers
}

\author{%
Pengfei~Zhang, \textit{Member, IEEE}, Xin~Xu, Hanrui~Wang, Xing~Xu, Bohao~Duan, Feng~Wei, \textit{Senior Member, IEEE}, Hongwei~Pan, \\
Xishui~Tian, Yi~Zhang, Pengxiong~Ma, and Olivier~Martineau\mbox{-}Huynh
\thanks{This research work is  supported by the following grants: the National SKA Program of China (2025SKA0110101)}
\thanks{Corresponding authors: Xin Xu (xxin\_1@stu.xidian.edu.cn)}%
\thanks{This is the accepted version of the article: P. Zhang \textit{et al.}, ``Noise Suppression and Radio Frequency Interference Rejection for Self-Triggered Radio Detectors of Extensive Air Showers,'' \textit{IEEE Transactions on Instrumentation and Measurement}, vol. 75, pp. 1--19, 2026, doi: 10.1109/TIM.2026.3677983. The final published version is available at IEEE Xplore. \copyright~2026 IEEE.}%
\thanks{Pengfei Zhang, Xin Xu, Hanrui Wang, Feng Wei, and Hongwei Pan are with the School of Electronic Engineering, Xidian University, Xi'an 710071, China.}%
\thanks{Xing Xu, Bohao Duan, Yi Zhang, and Pengxiong Ma are with the Purple Mountain Observatory, Chinese Academy of Sciences, Nanjing 210023, China.}%
\thanks{Xishui Tian is with the Department of Astronomy, School of Physics, Peking University, Beijing 100871, China.}%
\thanks{Olivier Martineau\mbox{-}Huynh is with LPNHE, Sorbonne Universit\'e, CNRS/IN2P3, 4~Pl. Jussieu, 75005 Paris, France, and also with Institut d'Astrophysique de Paris (CNRS UMR 7095), Sorbonne Universit\'e, 98~bis bd Arago, 75014 Paris, France.}%
}

\maketitle

\begin{abstract}

Self-triggered radio detection of ultra-high-energy cosmic rays and neutrinos offers a scalable and cost-effective approach for next-generation astroparticle observatories, but remains challenging under realistic radio-frequency interference (RFI) conditions. In the classical air-shower radio band, the achievable sensitivity and trigger reliability are critically limited by the balance between external sky background noise and internal detector-unit noise, as well as by non-stationary anthropogenic interference.

In this work, we present an end-to-end design and experimental characterization of a self-triggered radio detector unit explicitly optimized to operate in a galactic-noise-dominated regime. Rather than focusing on a single hardware component or trigger algorithm, we adopt a system-level methodology that coherently integrates sky-noise modeling, RF-chain noise budgeting, electromagnetic compatibility (EMC) mitigation, and measurement-driven validation.

By using the galactic radio background as a quantitative reference, we assess the internal noise performance of the detector unit and demonstrate conditions under which extensive air shower (EAS) radio signals can be distinguished from anthropogenic interference at the system-response level.

We further introduce an indirect noise-quantification method to estimate the low-noise amplifier contribution within the complete RF chain based on differential internal-noise measurements evaluated at the ADC level.

The proposed detector unit is validated through laboratory and on-site measurements, demonstrating operation close to the galactic-noise limit in the core frequency band. These results provide a practical and transferable methodology for the design and deployment of large-scale self-triggered radio arrays such as GRAND.

\end{abstract}


\section{Introduction}

\IEEEPARstart{T}{he} origin and acceleration mechanisms of Ultra-High-Energy Cosmic Rays (UHECRs, $E > 10^{15}\,\text{eV}$) and neutrinos remain among the major unsolved problems in modern astrophysics. These particles probe extreme astrophysical environments and provide unique tests of fundamental physics at unprecedented energy scales~\cite{tan2014}. Their detection therefore represents a central topic in contemporary astroparticle physics.

Extensive air showers (EAS) initiated by UHECRs in the atmosphere produce several observable signatures, including secondary particle cascades, nitrogen fluorescence emission, and coherent electromagnetic radiation. These signals are exploited by complementary detection techniques. Particle detector arrays based on scintillators or water Cherenkov detectors (e.g., Telescope Array, Auger, and the KM2A sub-array of LHAASO) measure secondary particles at ground level. Fluorescence detectors (e.g., Auger and TA) provide nearly calorimetric measurements of the shower development with excellent energy and composition resolution, but suffer from a limited duty cycle of about $10\%$. In contrast, radio arrays such as LOFAR, Auger, and GRAND offer nearly full-sky coverage with a duty cycle exceeding $90\%$, competitive energy resolution, and relatively low deployment cost~\cite{mayotte2025,abu2012,webster2021,almeida2022,martineau2025}. The SKA-LOW array has also initiated cosmic-ray observations, further extending the capabilities of low-frequency radio detection~\cite{wayth2021}.

Radio detection of cosmic rays was already explored in the mid-20th century~\cite{allan1971, bolton1948}, but only became experimentally viable in the early 21st century owing to advances in low-noise electronics, high-speed digitization, and large-scale data transmission. These developments have enabled modern radio arrays to achieve precise reconstruction of shower parameters and have established radio detection as a mature technique for UHECR studies~\cite{bolton1951}.

Neutrinos constitute an equally powerful astrophysical messenger. Their neutral charge allows them to point back to their sources, while their extremely small interaction cross-sections enable them to probe dense environments and cosmological distances~\cite{seckel2005}. In particular, the detection of cosmogenic neutrinos can provide critical tests of the physical processes associated with the Greisen--Zatsepin--Kuzmin (GZK) cutoff~\cite{kotera2010,heinze2016}. Among various detection strategies, Earth-skimming $\nu_\tau$ interactions are of special interest: a tau neutrino interacting below the Earth's surface can produce an emerging $\tau$ lepton that decays in the atmosphere and initiates an EAS. The resulting coherent radio emission in the $30$--$100\,\text{MHz}$ band can be detected using sparse radio arrays, forming the core detection concept of the GRAND project~\cite{alvarez2020}.

Achieving significant statistics in the ultra-high-energy regime requires extremely large detection areas due to the low fluxes of UHECRs and neutrinos. 
This requirement is further amplified by the steeply falling energy spectra of cosmic rays and neutrinos, which approximately scale as $E^{-3}$ and $E^{-2}$, respectively, implying that even a modest reduction of the effective trigger threshold can lead to a substantial increase in detectable event rates.
Sparse, self-triggered radio arrays provide a technically and economically viable solution to this challenge~\cite{charrier2019,barwick2017}. However, self-triggered operation in the classical radio detection band ($30$--$350\,\text{MHz}$) exposes detector units to substantial noise and radio-frequency interference (RFI). In addition to the diffuse sky background, anthropogenic sources such as broadcasting transmitters, communication systems, power infrastructure, and radar installations introduce both transient and continuous-wave interference.

These constraints are particularly severe for fully self-triggered radio arrays.
Unlike externally triggered experiments such as the Pierre Auger Observatory, where raw waveforms from radio antennas can be filtered offline, a self-triggered system must make real-time decisions based solely on local measurements.
As a consequence, elevated noise levels or high rates of transient interference can render the detector effectively blind to weak EAS signals.

Traditional electromagnetic compatibility (EMC) practices to address these challenges include hardware-level shielding and grounding~\cite{danker2004}, RF-chain optimization and low-noise amplifier design~\cite{chiong2021}, analog filtering to suppress out-of-band interference~\cite{cameron2018}, and digital signal-processing techniques to mitigate persistent narrowband signals~\cite{offringa2010}. In addition, careful selection of radio-quiet deployment sites is commonly employed. Nevertheless, these approaches are often applied in a fragmented manner and must be adapted to the specific architecture and environment of each detector system.

In this work, we present both a systematic, end-to-end design of a self-triggered radio detector unit, and a protocol to 
characterize and optimize it to reach performances allowing to operate it in a regime dominated by galactic noise. By coherently combining sky-background modeling, RF-chain noise budgeting, electromagnetic compatibility (EMC) design, and measurement-driven validation, this work provides, to the best of our knowledge, the first comprehensive and quantitative detector-unit-level review of noise sources and RF-chain architecture for self-triggered extensive air-shower radio experiments.

At the system level, the main innovations and contributions of this work can be summarized as follows.
First, we establish a quantitative, sky-noise-referenced benchmark for self-triggered radio detector units operating in the classical 30--350~MHz band. By explicitly modeling the Galactic radio emission and propagating it through the complete RF chain to the ADC level, a physically motivated target noise level is defined, providing a unified reference for detector-unit design and performance validation.
Second, we develop an end-to-end detector-unit noise-control methodology that jointly integrates RF-chain optimization, electromagnetic compatibility (EMC) design, and measurement-driven validation. This approach systematically identifies dominant noise sources and coupling paths within the integrated system, rather than addressing individual components in isolation.
Third, we provide a system-level physical basis for mitigating anthropogenic radio-frequency interference in self-triggered operation. Continuous-wave interference is suppressed through dedicated analog and digital filtering, while transient anthropogenic signals are characterized using waveform-level temporal and spectral features, thereby reducing false triggers while preserving the broadband impulsive response required for EAS radio detection.

To make the notion of the ``target noise level'' explicit, we emphasize that it is defined with respect to the irreducible Galactic sky background received by an omnidirectional antenna system.
Unlike directional radio detectors, which observe only a limited solid angle and can partially suppress the Galactic contribution, the tri-polarized, quasi-omnidirectional antennas adopted by GRAND inevitably collect radio emission from nearly the entire upper hemisphere.
As a consequence, the Galactic background constitutes a fundamental and non-removable noise floor for the detector units.

The structure of the paper is as follows. Section~II introduces the working principle and standard implementation of the self-triggered array. Section~III establishes the sky background noise analysis and the corresponding system response. Section~IV details the electromagnetic compatibility (EMC) implementation. Section~V validates the noise suppression strategy. Section~VI discusses external pulse interference sources. Section~VII concludes the paper.
The present work is carried out within the framework of the GRAND collaboration, and is validated using the GRANDProto300 prototype deployed in the Gobi Desert.

\section{Introduction to the Self-trigger Radio Detection System and the Challenges of Noise/Interference Processing}

\subsection{Working Principle of Self-trigger Radio Detection System and Typical Unit Composition}

\begin{figure*}[ht]
  \centering
  \includegraphics[width=7.16in]{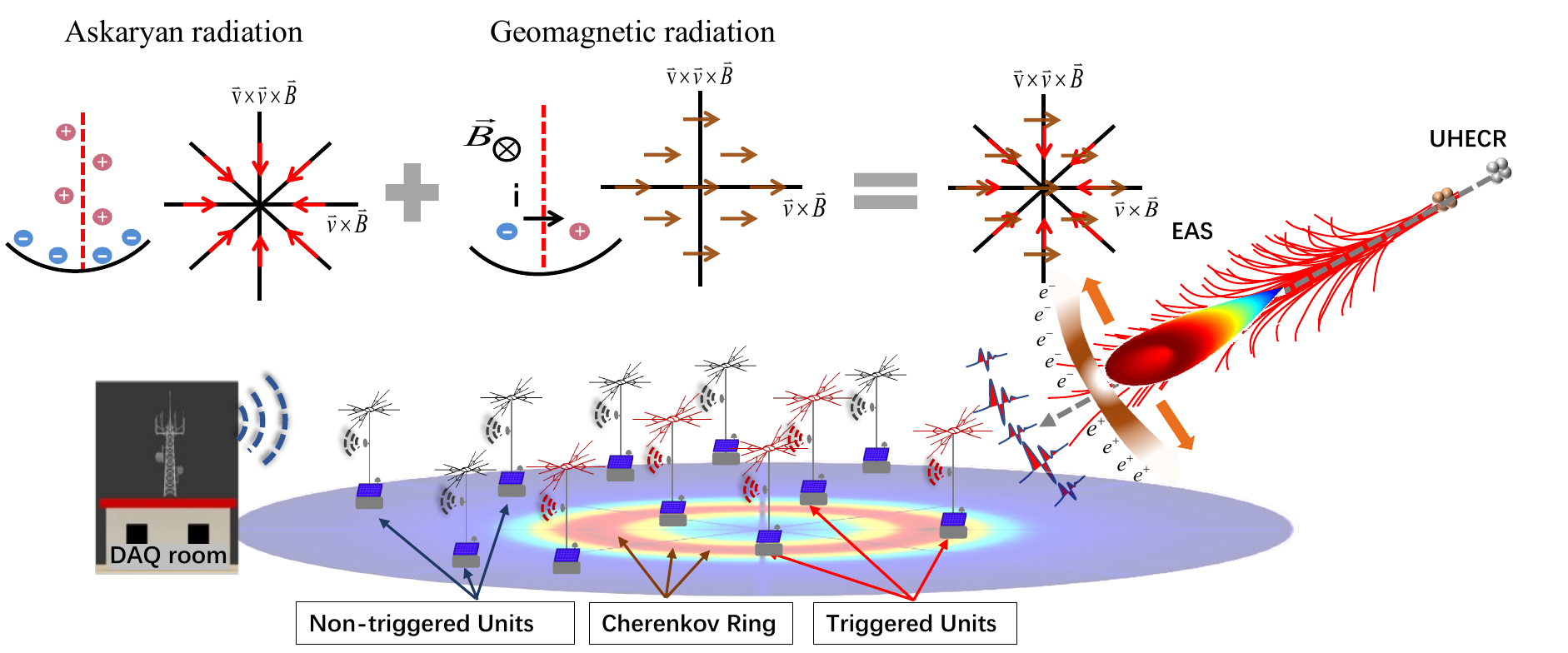}
  \caption{Principle of high-energy cosmic ray extensive air shower (EAS) detection based on a self-triggered low-frequency radio sparse array.}
  \label{fig:principle}
\end{figure*}

Fig.~\ref{fig:principle} illustrates the working principle of a typical self-triggered radio detection array. When a high-energy cosmic ray enters the Earth's atmosphere, it initiates an extensive air shower (EAS) composed of a large number of secondary charged particles~\cite{kampert2012,cummings2021}. During the shower development, a negative charge excess builds up in the shower front, giving rise to Askaryan radiation through coherent Cherenkov emission in dense media~\cite{askaryan1962}. In addition, the deflection of charged particles in the Earth's magnetic field produces geomagnetic radiation~\cite{kahn1966}. The radio emission remains coherent for wavelengths significantly larger than the shower pancake thickness ($d \sim 1$--$3\,\text{m}$), resulting in macroscopic radiation strength at frequencies below $\sim 100\,\text{MHz}$~\cite{alvarez2012}. Atmospheric propagation with refractive index $n(h)>1$ further leads to the Cherenkov compression effect, producing short impulsive pulses with durations below $10\,\text{ns}$ and frequency components extending up to $200\,\text{MHz}$ and beyond around the Cherenkov angle $\theta_c \approx 0.5^\circ$--$1^\circ$~\cite{huege2013}. Once the radio signal reaches the ground, it may trigger individual antenna stations if its amplitude exceeds a predefined threshold. Triggered detection units transmit digitized waveforms and precise timing information to a central data acquisition (DAQ) system, where temporal coincidences are used to reconstruct the EAS event. Subsequent offline processing deconvolves the antenna response~\cite{zhang2025} and reconstructs the energy, mass composition, and arrival direction of the primary particle~\cite{guelfand2025,schluter2023}.

The core functionalities of a self-triggered radio detection unit include radio-frequency pulse acquisition and digitization, precise timing via GPS, data transmission, autonomous power supply, and monitoring of auxiliary parameters.

\begin{figure}[!t]
  \centering
  \includegraphics[width=3.0in]{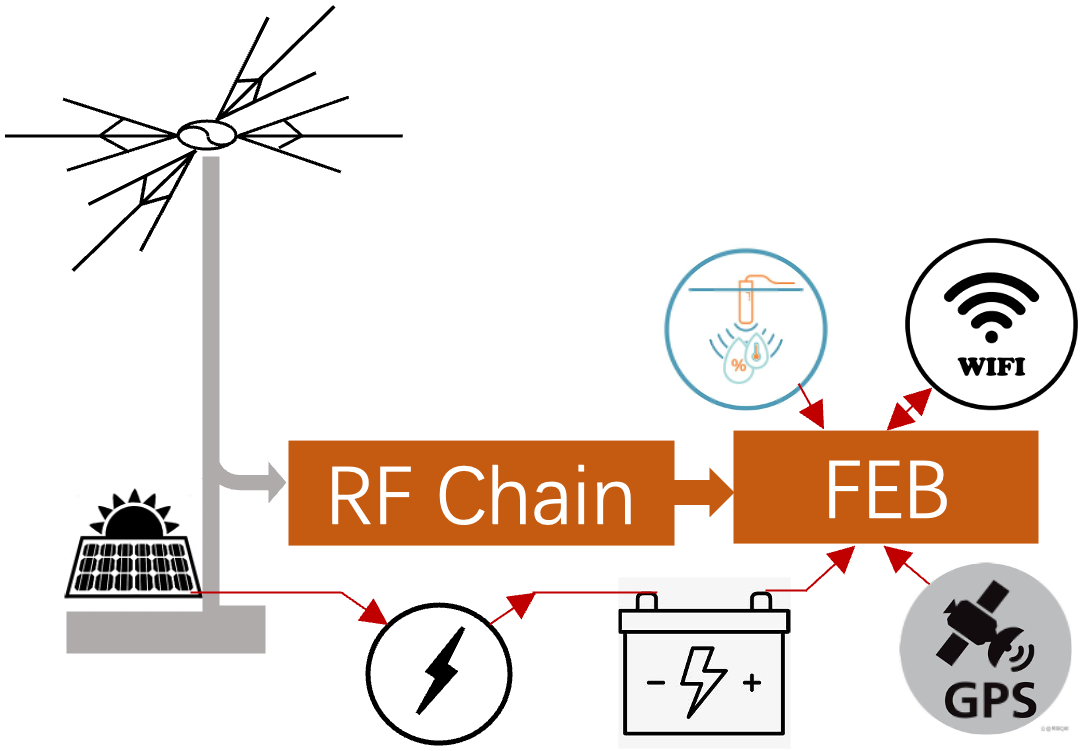}\\[1ex]
  (a)\\[2ex]
  \includegraphics[width=3.0in]{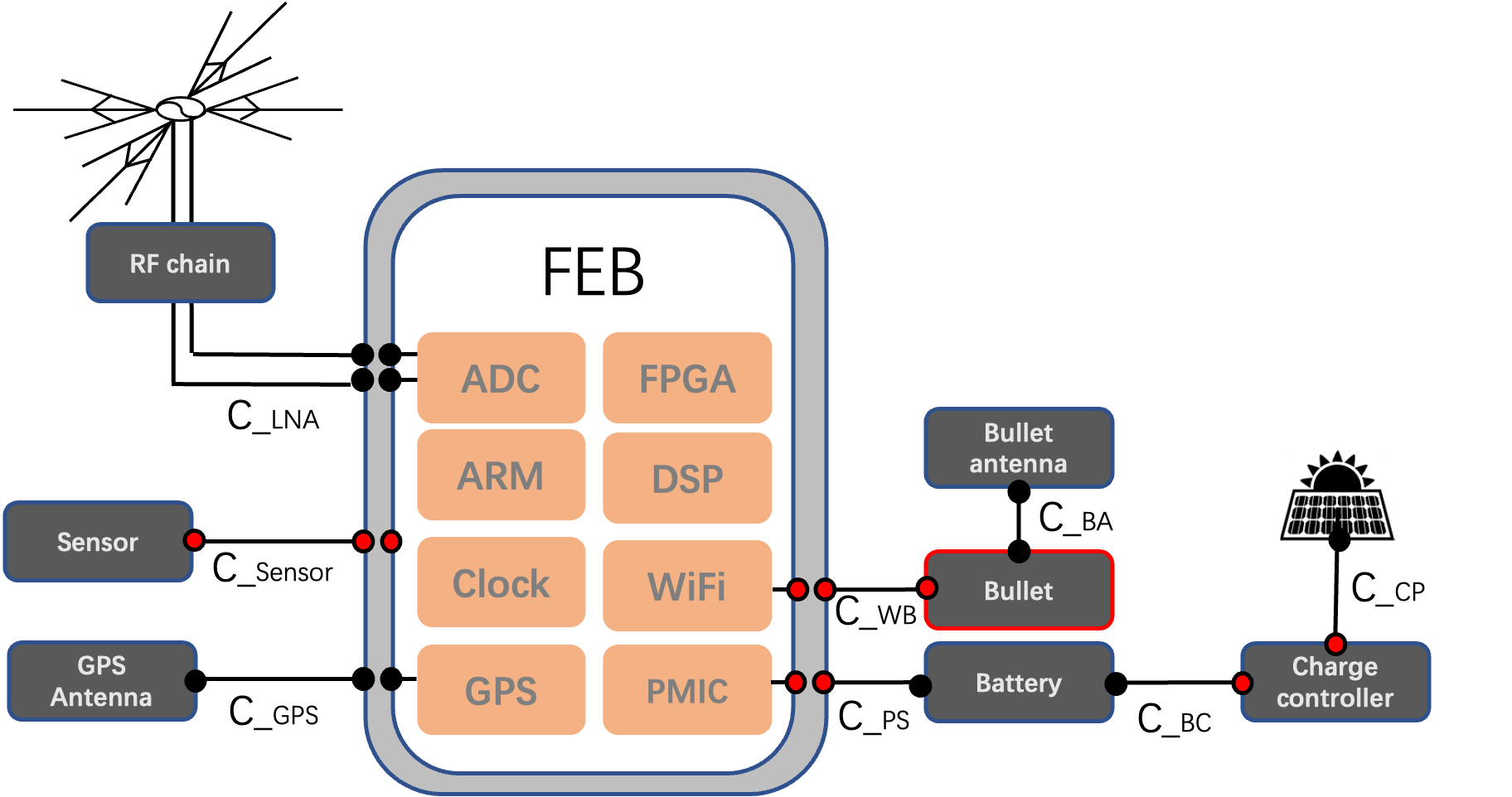}\\[1ex]
  (b)
  \caption{Illustration of a typical detection unit: (a) functional module connection diagram; (b) cable connection topology diagram.}
  \label{fig:unit1}
\end{figure}

Fig.~\ref{fig:unit1}(a) shows a representative architecture of a self-triggered detection unit. Solar panels harvest energy and recharge a battery through a charge controller, providing autonomous power to the system. The battery supplies the front-end board (FEB), which serves as the central hub connecting the antenna and RF chain, timing and sensor subsystems, and the communication module linked to the central DAQ system.

As illustrated in Fig.~\ref{fig:unit1}(b), the FEB interconnects all functional components via dedicated cabling. These interconnections can unintentionally act as electromagnetic interference (EMI) coupling paths: noise generated by internal subsystems may couple onto cables, radiate into the surrounding environment, and be re-captured by the antennas, forming an internal feedback loop that contaminates the received signal. This coupling mechanism highlights the importance of rigorous electromagnetic compatibility (EMC) design for both system components and their interconnections.

\subsection{Typical RFI and Noise in Low Frequency Detection system}

\begin{figure*}[ht]
  \centering
  \includegraphics[width=7.16in]{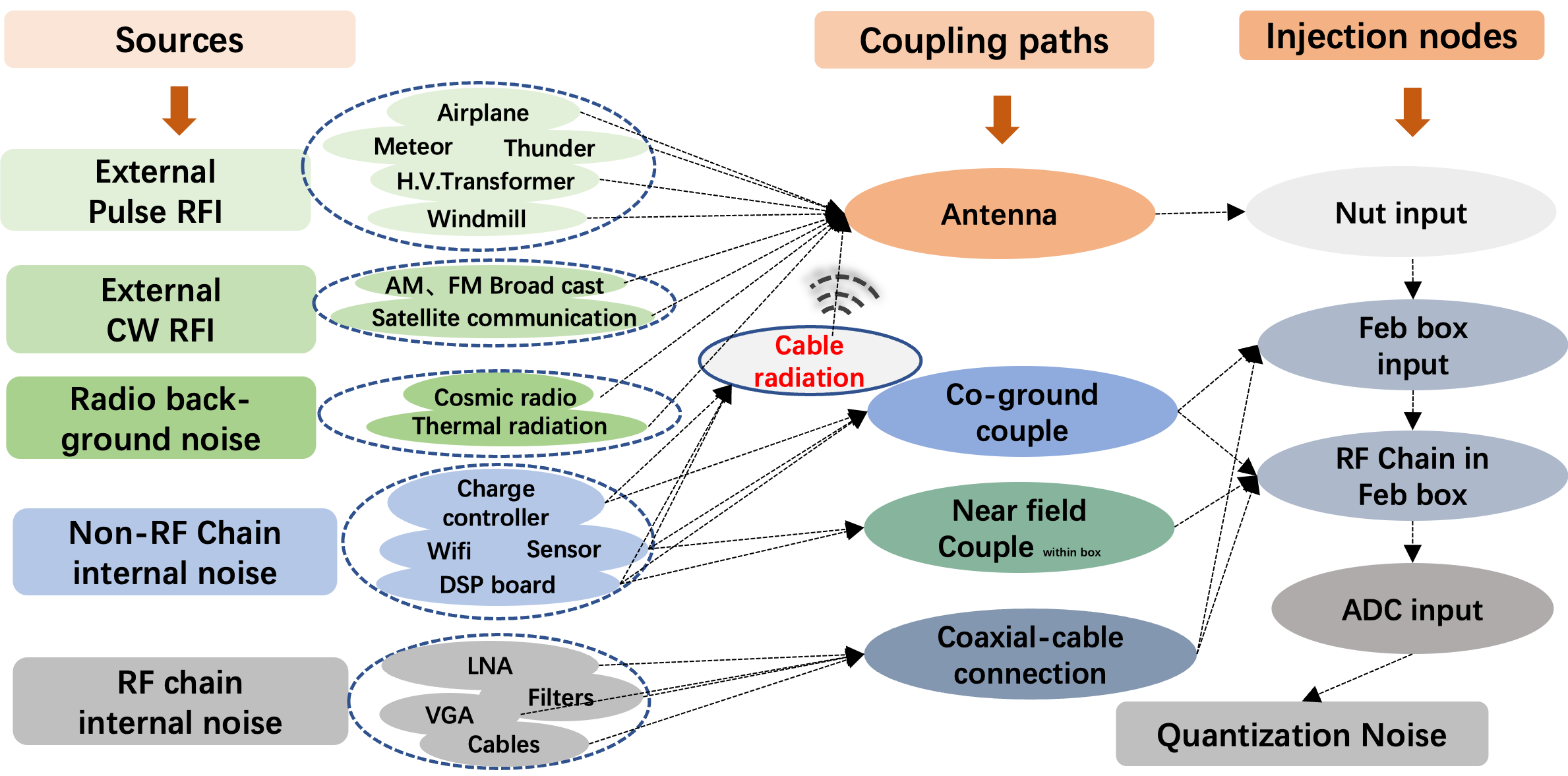}
  \caption{Schematic diagram of the interference/noise sources, coupling methods, and injection nodes faced by the GP300 unit}
  \label{fig:Schematic diagram}
\end{figure*}

Low-frequency self-triggered radio detection systems are subject to a complex electromagnetic environment composed of both interference and intrinsic noise. Fig.~\ref{fig:Schematic diagram} provides an overview of the dominant noise and interference sources, their coupling mechanisms, and their injection points along the analog front-end signal chain. For clarity, interference is classified into transient pulse interference and continuous-wave (CW) interference, while noise contributions include radio background noise, instrumental noise, RF-chain noise, and ADC quantization noise. This classification forms the basis for the mitigation strategies discussed in subsequent sections.

\subsubsection{External Pulse Interference}

External transient electromagnetic pulses generated by anthropogenic or natural sources may exhibit temporal, spectral, and polarization characteristics comparable to those of extensive air shower (EAS) radio signals. These transients therefore constitute a major source of false triggers in self-triggered detection systems. Although offline analyses exploiting event topology and multi-station correlations can reject a significant fraction of such events, realistic operating environments generally require online mitigation to limit data volume and preserve system live time.

\subsubsection{External Continuous-Wave Interference}

Continuous-wave interference consists of persistent narrowband signals originating from broadcasting services and satellite communications. Such signals may be present on time scales ranging from minutes to continuous operation. When sufficiently strong, CW interference can saturate receiver stages or suppress the triggering of weak broadband EAS pulses, making dedicated analog and digital mitigation strategies essential.

\subsubsection{Instrumental Noise (Non-RF Chain)}

Instrumental noise originates from auxiliary electronic subsystems that are not part of the RF signal chain, including power regulation modules, communication interfaces, environmental sensors, and digital control circuits. Although these subsystems do not directly process radio signals, their electromagnetic emissions can couple into the RF chain through cable radiation, ground loops, and near-field coupling, thereby contaminating the received signal.

\subsubsection{RF-Chain Noise of the Observation System}

Noise generated within the RF chain arises from the intrinsic electronic noise of components such as baluns, low-noise amplifiers (LNAs), filters, variable-gain amplifiers (VGAs), and ADC interfaces. Since the sky background noise defines the ultimate sensitivity limit, RF-chain design and noise-suppression strategies aim to reduce all internal noise contributions below the sky-noise level within the core operational band.

\subsection{Challenges of Self-Triggered-Based Weak Pulse Detection}

Self-triggered radio detection imposes stringent requirements on noise suppression and signal discrimination because of the intrinsic properties of radio pulses emitted by extensive air showers (EAS). These pulses are characterized by low amplitudes, short durations, and broadband frequency content, which collectively result in weak signal-to-noise ratios (SNRs) at the antenna terminals. Consequently, achieving high trigger efficiency without an excessive false-trigger rate represents a central technical challenge for self-triggered systems.

The electric-field amplitudes of EAS-induced radio signals are typically below $100\,\mu\text{V}\,\text{m}^{-1}$ at ground level. After projection onto a limited number of fixed antenna polarizations and subsequent analog processing, the effective voltage amplitudes at the receiver input are further reduced. At the same time, the intrinsic noise power of each polarization channel remains unchanged, leading to a reduction of the effective per-channel SNR.

In the time domain, EAS radio pulses exhibit durations on the order of a few to several tens of nanoseconds, depending on the observer position relative to the Cherenkov ring. Such short impulses correspond to broadband frequency spectra extending from tens of megahertz up to the gigahertz range. In practice, the usable bandwidth is limited by the antenna response and the analog front-end, further constraining the recoverable signal power.

In addition, the polarization of the radio emission is determined by the vector superposition of geomagnetic and Askaryan contributions. When measured with single- or dual-polarization antennas, the incident electric field is decomposed onto a limited number of fixed polarization axes. This projection distributes the signal power among channels while leaving the noise level unchanged, thereby further reducing the trigger probability compared to an ideal full-vector measurement.

As a result of these combined effects, self-triggered detection requires highly sensitive weak-pulse triggering capabilities. Figure~\ref{fig:Schematic diagram} illustrates the various noise and interference sources contributing to the total system noise. These contributions can be grouped into three categories: fundamental noise sources, reducible internal noise, and suppressible external interference.

Fundamental noise sources include the sky background and thermal noise. Although the sky background contribution can, in principle, be reduced by restricting the antenna response to directions near the horizon, such configurations would significantly limit the accessible solid angle and thus the detection efficiency, especially for small or sparse arrays. To preserve a wide field of view, the sky background noise must therefore be tolerated as an irreducible noise floor.

Reducible internal noise sources include RF-chain noise and non-RF instrumental noise, both of which can be mitigated through optimized electronic design, electromagnetic compatibility measures, and careful system integration. Suppressible external interference comprises continuous-wave emissions and transient anthropogenic pulses, which can be addressed through a combination of analog filtering, digital processing, and trigger-level discrimination algorithms.

For a given antenna design, the sky background ultimately defines the lowest achievable system noise level. The primary design objective of the self-triggered detector presented in this work is therefore to suppress all reducible noise contributions--namely RF-chain noise, instrumental noise, and external CW interference--below the sky background level within the core operational band. Section~III quantifies this target using sky-noise modeling and system response analysis, while Section~IV describes the corresponding implementation strategies.

\section{Analysis of RF Chain Response and Sky Background Noise}

Sky background noise provides a well-defined and irreducible reference for evaluating the sensitivity of low-frequency radio detection systems. By leveraging precise sky temperature models together with measured antenna and RF-chain responses, the detector response to cosmic radio background can be accurately quantified. This section introduces the architecture of a typical RF chain, establishes a formal description of its key parameters, and derives the expected system response to sky background noise, which serves as a benchmark for subsequent noise-suppression design.

\subsection{Architecture and Parameter Representation of a Typical RF Chain}

\begin{figure}[!t]
  \centering
  \includegraphics[width=3.4in]{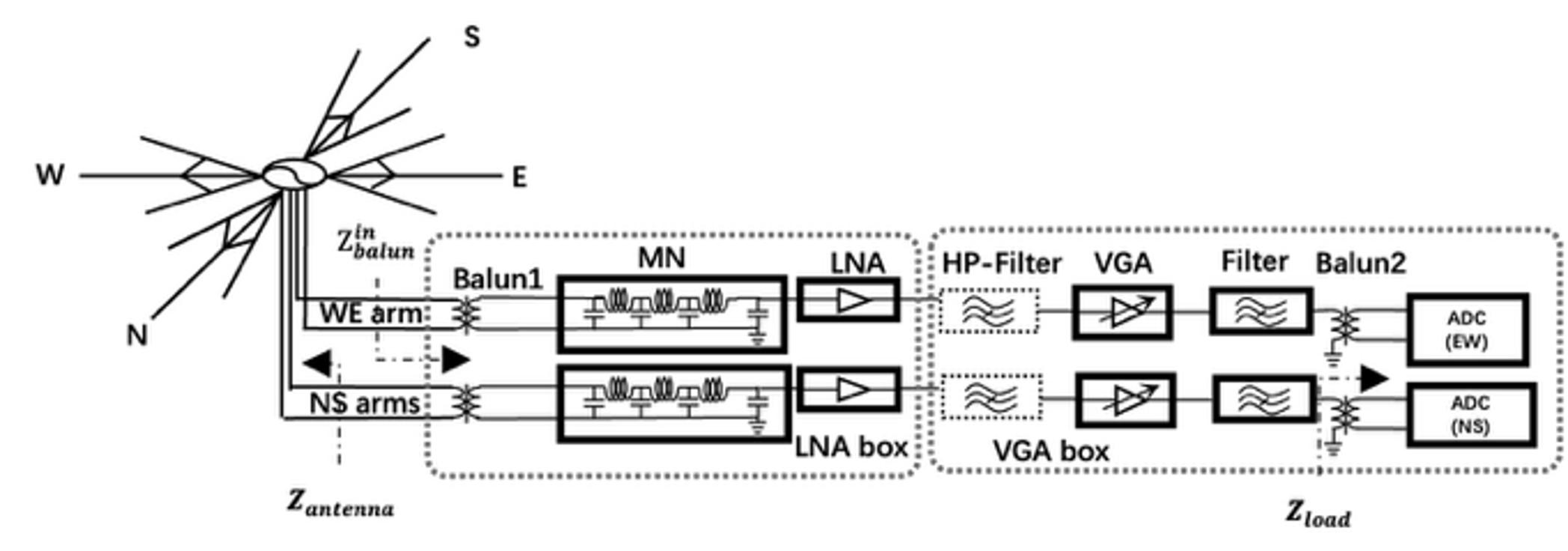}\\[1ex]
  (a)\\[2ex]
  \includegraphics[width=3.4in]{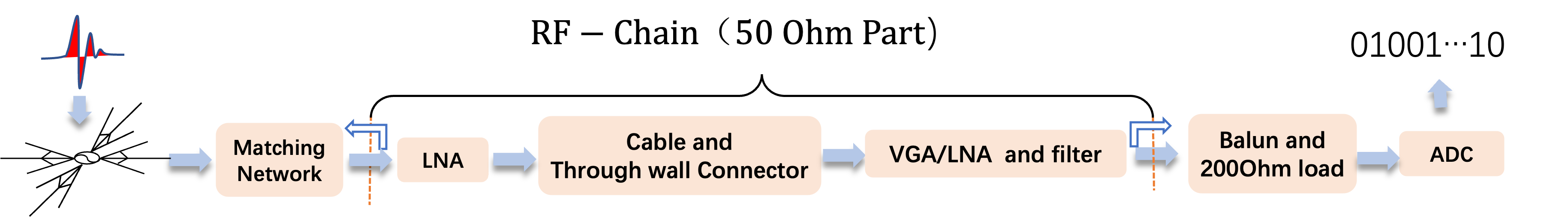}\\[1ex]
  (b)
  \caption{Schematic of the GRANDProto RF Chain Architecture. (a) Electrical connectivity diagram (Note: Early designs omitted HP-Filter for AM suppression; modular implementation allows post-installation) (b) Physical module configuration }
  \label{fig:unit2}
\end{figure}

Fig.~\ref{fig:unit2} illustrates the RF-chain architecture of the GRANDProto detector unit~\cite{grandproto}, which is used here as a representative example to demonstrate a generalizable analysis method. During operation, the antenna signal is converted from differential to single-ended mode by a balun, matched to the low-noise amplifier (LNA) through a matching network, and subsequently amplified. If required, a high-pass filter (HPF) for AM-band suppression can be inserted after the LNA to protect downstream components~\cite{batista2025}. The amplified signal is then transmitted via an RF cable to the variable-gain amplifier (VGA), followed by band-limiting filters to suppress out-of-band noise and prevent aliasing before digitization. Finally, the signal is converted back to differential mode and sampled by the ADC. This modular architecture facilitates flexible deployment while limiting self-coupled noise pickup by the antenna.

The RF system is characterized by three groups of parameters: (i) the antenna impedance $\mathbf{Z}_\text{antenna}^{SN/EW/V}$ and its equivalent length matrix $\mathbf{l}$, (ii) the RF-chain transmission matrix $\mathbf{A}$, and (iii) the ADC load impedance $Z_\text{load}$. This formalism enables a compact and consistent description of the full signal path from the incident electric field to the digitized voltage.

The equivalent length matrix of the three-channel antenna system can be expressed as
\begin{equation}
\mathbf{l} =
\begin{bmatrix}
l_{SN}^{SN} & l_{EW}^{SN} & l_{V}^{SN} \\
l_{SN}^{EW} & l_{EW}^{EW} & l_{V}^{EW} \\
l_{SN}^{V}  & l_{EW}^{V}  & l_{V}^{V}
\end{bmatrix}.
\label{eq:lmatrix}
\end{equation}

Similarly, the RF-chain transmission matrix is written as
\begin{equation}
\mathbf{A} =
\begin{bmatrix}
A^{RF-Chain}_{11} & A^{RF-Chain}_{12} \\
A^{RF-Chain}_{21} & A^{RF-Chain}_{22}
\end{bmatrix}.
\label{eq:Amatrix}
\end{equation}

The antenna parameters can be obtained from EM simulations or direct measurements. 
The RF-chain parameters are determined by measuring the scattering parameters $S_{ij}$ 
of each component using a vector network analyzer (VNA). 

These scattering parameters are then converted into the transmission-matrix elements $A_{ij}$ 
via the standard relations:
\begin{equation}
\scriptsize
\mathbf{A} = 
\frac{1}{2s_{21}}
\begin{bmatrix}
(1+s_{11})(1-s_{22})+s_{12}s_{21} & (1+s_{11})(1+s_{22})-s_{12}s_{21} \\
(1-s_{11})(1-s_{22})-s_{12}s_{21} & (1-s_{11})(1+s_{22})+s_{12}s_{21}
\end{bmatrix}
\normalsize
\label{eq:S2A}
\end{equation}

The advantage of using the transmission-matrix formalism is that the overall response 
of a cascaded RF chain can be obtained by simple matrix multiplication of the individual 
component matrices, i.e.,
\begin{equation}
\scriptsize
\mathbf{A}^{RF-Chain} = 
\mathbf{A}^{Balun1}\,
\mathbf{A}^{MN}\,
\mathbf{A}^{LNA}\,
\mathbf{A}^{Cable}\,
\mathbf{A}^{VGA}\,
\mathbf{A}^{Filter}\,
\mathbf{A}^{Balun2}.
\label{eq:cascade}
\end{equation}

Marking Balun1, matching network, LNA, cable, VGA, filter, and Balun2 as devices $i=1\dots 7$ in sequence, the cascaded formula can also be expressed as
\begin{equation}
\mathbf{A}^{RF-Chain} = \prod_{i=1}^{7} \mathbf{A}^i.
\label{eq:product}
\end{equation}

\subsection{Response of the Antenna and RF Chain to an Incident Plane Wave}

For an incident plane wave with electric field vector $\vec{E}_{in}$, the antenna response can be described using the equivalent length formalism. The open-circuit voltages induced at the antenna terminals are obtained through the projection of the incident field onto the antenna equivalent length matrix.

\begin{equation}
\begin{bmatrix}
V^{SN}_{OC} \\
V^{EW}_{OC} \\
V^{V}_{OC}
\end{bmatrix}
=
\begin{bmatrix}
l^{SN}_{SN} & l^{SN}_{EW} & l^{SN}_{V} \\
l^{EW}_{SN} & l^{EW}_{EW} & l^{EW}_{V} \\
l^{V}_{SN}  & l^{V}_{EW}  & l^{V}_{V}
\end{bmatrix}
\begin{bmatrix}
E^{SN}_{in} \\
E^{EW}_{in} \\
E^{V}_{in}
\end{bmatrix}.
\label{eq:Voc}
\end{equation}

Considering that the input impedance from the antenna to the balun is
\begin{equation}
Z^{in}_{Balun} =
\frac{A^{RF-Chain}_{11}Z_{load} + A^{RF-Chain}_{12}}
     {A^{RF-Chain}_{21}Z_{load} + A^{RF-Chain}_{22}},
\label{eq:Zin}
\end{equation}
the voltage and current at the input of the balun can be expressed as
\begin{equation}
V^{in}_{Balun} = V_{OC}\,
\frac{Z^{in}_{Balun}}{Z_{antenna}+Z^{in}_{Balun}},
\label{eq:Vin}
\end{equation}
\begin{equation}
I^{in}_{Balun} = 
\frac{V_{OC}}{Z_{antenna}+Z^{in}_{Balun}}.
\label{eq:Iin}
\end{equation}

Finally, the voltage at the ADC input $V^{out}_{RF-Chain}$ can be obtained 
through the following matrix relationship:
\begin{equation}
\begin{bmatrix}
V^{out}_{RF-Chain} \\
I^{out}_{RF-Chain}
\end{bmatrix}
=
\begin{bmatrix}
A^{RF-Chain}_{11} & A^{RF-Chain}_{12} \\
A^{RF-Chain}_{21} & A^{RF-Chain}_{22}
\end{bmatrix}^{-1}
\begin{bmatrix}
V^{in}_{Balun} \\
I^{in}_{Balun}
\end{bmatrix}.
\label{eq:Vout}
\end{equation}

\subsection{Calculation of Sky Background Noise Response}

While the previous formulation describes the response to a plane wave from a specific direction, the sky background noise requires integration over the full sky. In this case, the open-circuit voltage formulation is replaced by an equivalent representation based on the total sky noise power received by the antenna.

\begin{equation}
V^{Sky}_{OC} = 2\sqrt{P_{Sky}} \times R_{antenna},
\label{eq:VocSky}
\end{equation}
with
\begin{equation}
P_{Sky}(f,t) = \frac{1}{2}\int df \int_{\Omega} 
B(f,\alpha,\delta,t)\,A_e(f,\alpha,\delta)\, d\Omega,
\label{eq:Psky}
\end{equation}
where $d\Omega$ denotes the elementary solid angle, $\alpha$ and $\delta$ 
are the zenith and azimuth angles. $A_e$ is the effective aperture of the antenna, 
and it relates to its gain through
\begin{equation}
A_e = \frac{\lambda^2}{4\pi} G.
\label{eq:Ae}
\end{equation}

The power spectral density of the sky background is expressed as
\begin{equation}
B(f,\alpha,\delta,t) = \frac{2k}{c^2}\, f^2 T_{sky}(f,\alpha,\delta,t),
\label{eq:Bsky}
\end{equation}
where $k$ is the Boltzmann constant. 
The sky temperature can be modeled as
\begin{equation}
T_{sky}(f,\alpha,\delta) = T_{CMB} + T_{ISO}(f) + T_{Galaxy}(f,\alpha,\delta),
\label{eq:Tsky}
\end{equation}
with $T_{CMB} = 2.73\,\text{K}$ denoting the cosmic microwave background temperature, 
$T_{ISO}(f)$ the frequency-dependent isotropic component of the sky background, 
and $T_{Galaxy}$ the Galactic noise contribution. 

The diffuse Galactic component and its spectral behavior at decametric and metric wavelengths have been extensively characterized by sky surveys and modeling efforts~\cite{platania2003,cane1979}.

To account for Earth's rotation, the sky temperature distribution is transformed into local horizontal coordinates and integrated together with the antenna response and obstruction masks. This procedure yields time-dependent sky-noise spectra $P_{Sky}(f,t)$ and the corresponding open-circuit voltage $V^{Sky}_{OC}$.

\begin{figure*}[!t]
  \centering
  \subfloat[30~MHz]{\includegraphics[width=0.22\linewidth]{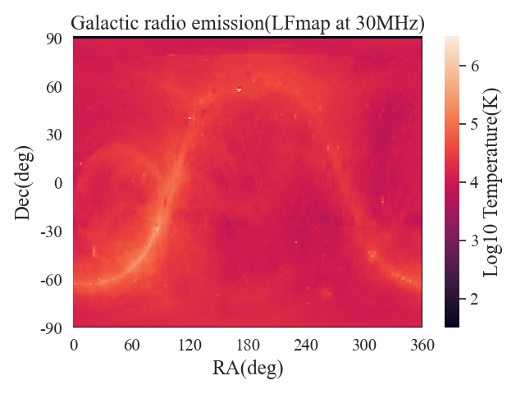}}
  \hfill
  \subfloat[60~MHz]{\includegraphics[width=0.22\linewidth]{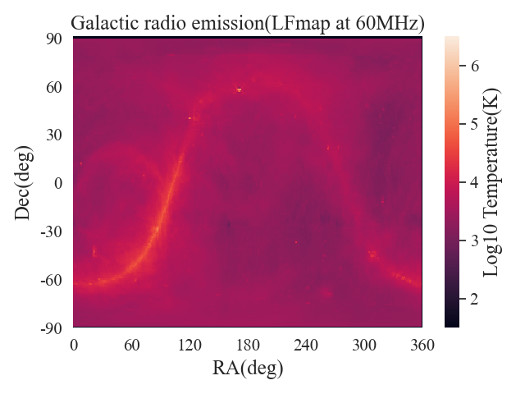}}
  \hfill
  \subfloat[90~MHz]{\includegraphics[width=0.22\linewidth]{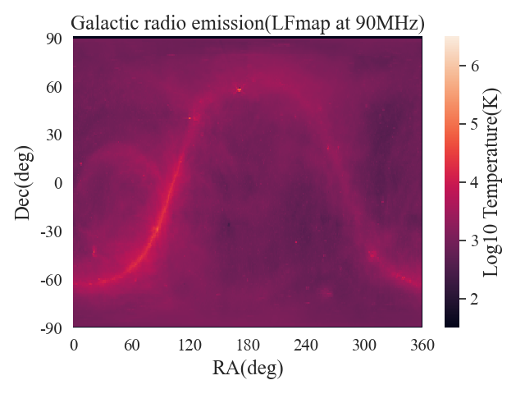}}
  \hfill
  \subfloat[120~MHz]{\includegraphics[width=0.22\linewidth]{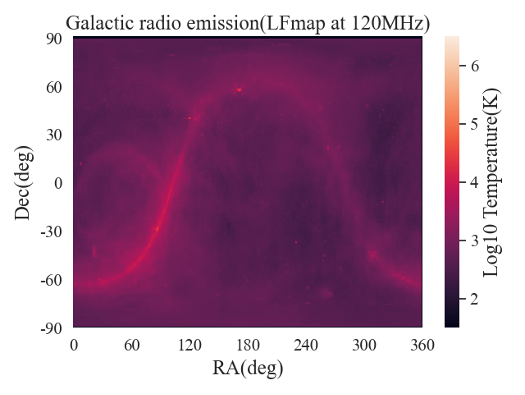}}\\[1ex]
  \subfloat[150~MHz]{\includegraphics[width=0.22\linewidth]{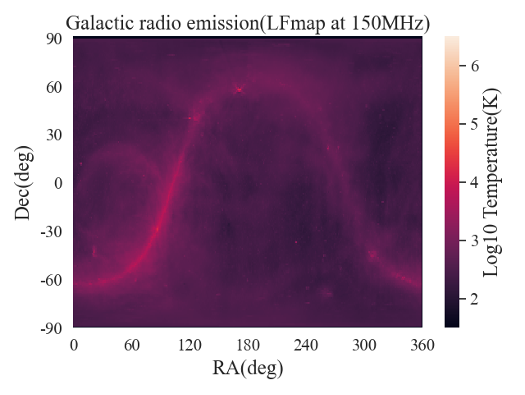}}
  \hfill
  \subfloat[180~MHz]{\includegraphics[width=0.22\linewidth]{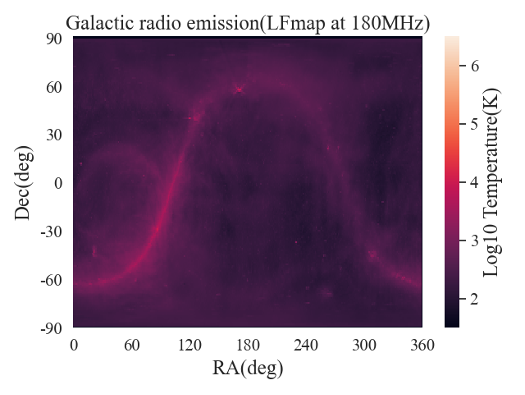}}
  \hfill
  \subfloat[210~MHz]{\includegraphics[width=0.22\linewidth]{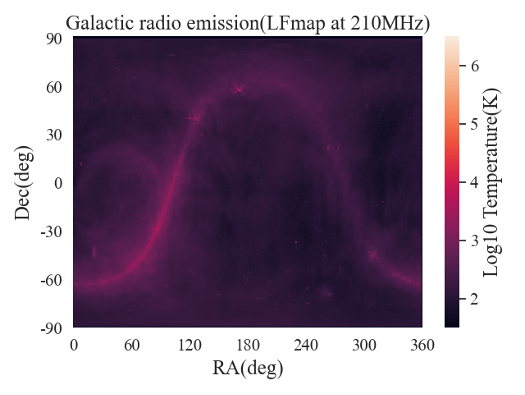}}
  \hfill
  \subfloat[250~MHz]{\includegraphics[width=0.22\linewidth]{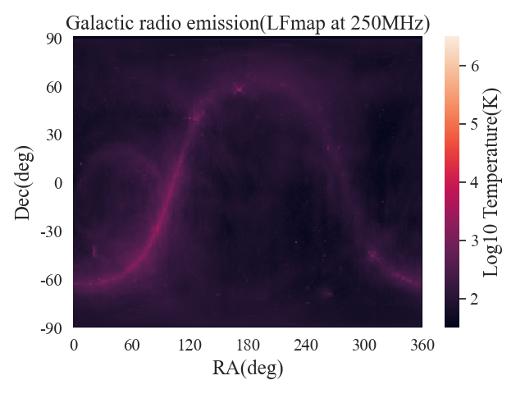}}\\[1ex]
  \subfloat[PSD, port~1]{\includegraphics[width=0.45\linewidth]{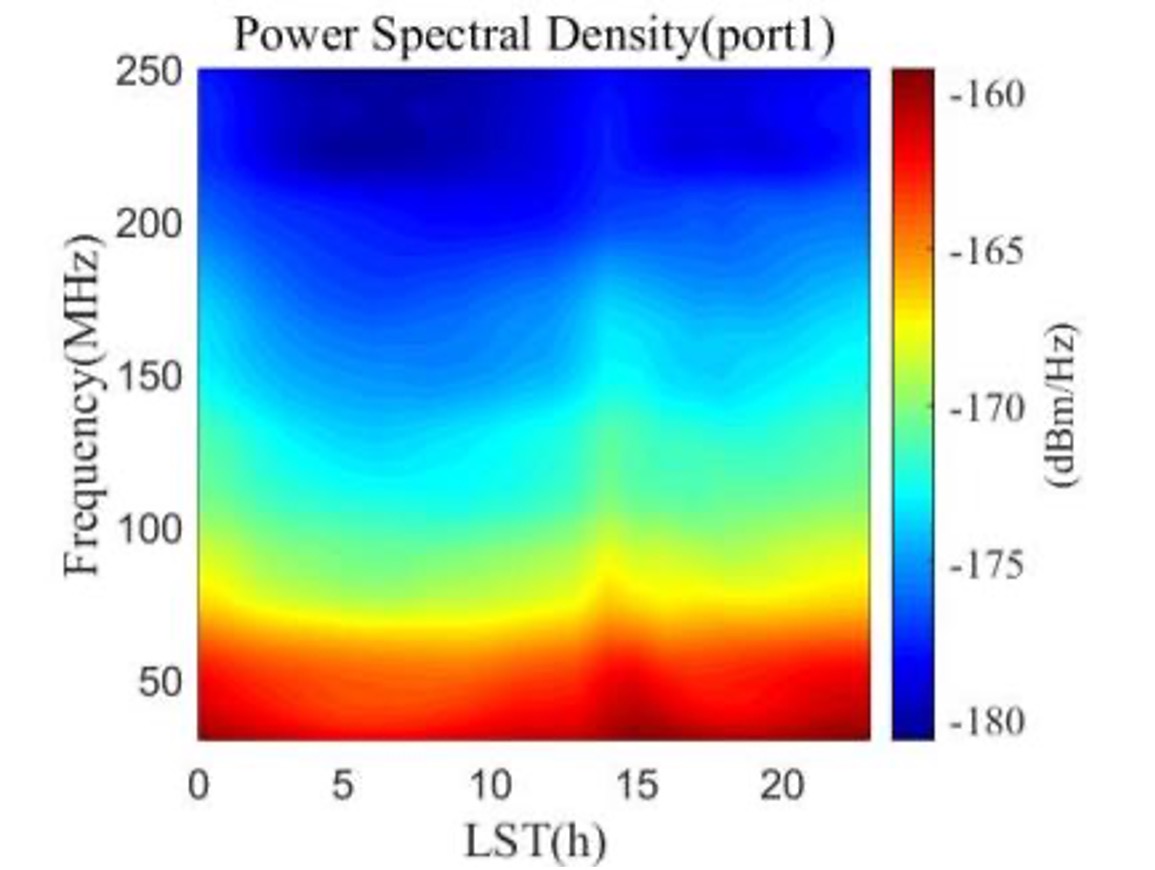}}
  \hfill
  \subfloat[PSD, port~2]{\includegraphics[width=0.45\linewidth]{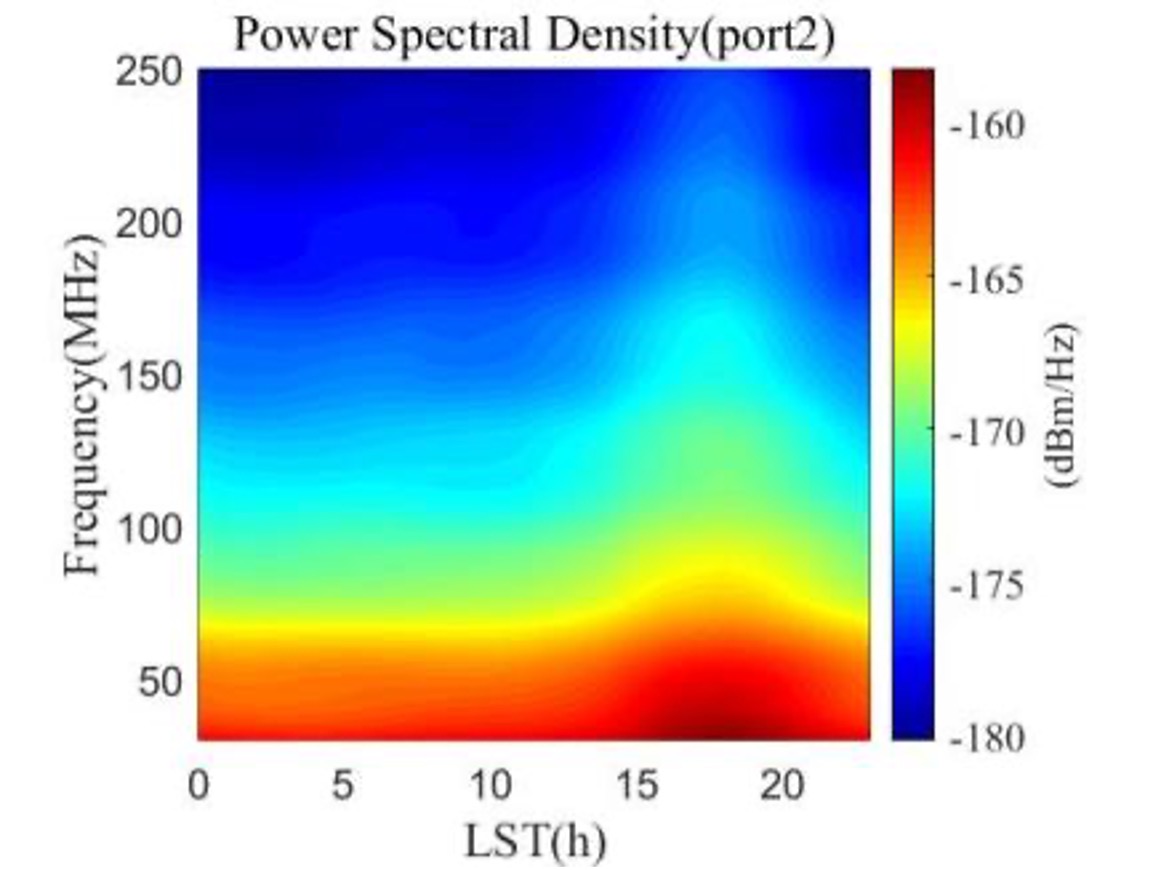}}
  \caption{Galactic noise brightness temperature maps and corresponding power spectral density (PSD). 
  (a)--(h) Sky temperature maps in equatorial coordinates based on LFmap at 30, 60, 90, 120, 150, 180, 210, and 250~MHz, respectively. 
  (i)--(j) Time-varying PSD obtained after transformation into local horizontal coordinates and convolution with the antenna response functions: (i) port~1, (j) port~2.}
  \label{fig:sky_noise}
\end{figure*}

Fig.~\ref{fig:sky_noise}(a) presents the modeled sky background brightness temperature in equatorial coordinates at representative frequencies between 30 and 250~MHz, based on LFmap~\cite{polisensky2007}. These sky maps are transformed into local horizontal coordinates and convolved with the full detector response, including the antenna pattern and the complete RF-chain transfer function. The resulting power spectral density (PSD) at the ADC input is shown in Fig.~\ref{fig:sky_noise}(b), providing a direct prediction of the sky-induced noise level at the system output.

\begin{figure}[!t]
  \centering
  \includegraphics[width=3.0in]{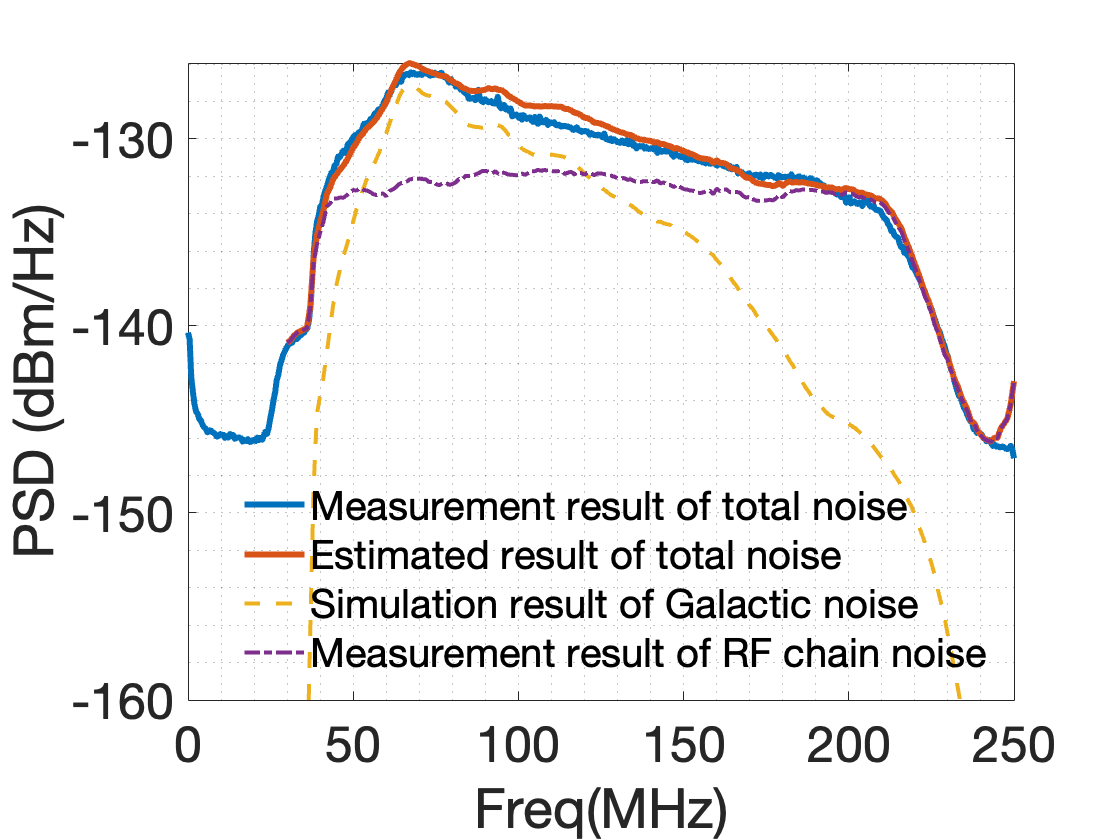}\\[1ex]
  (a)\\[2ex]
  \includegraphics[width=3.0in]{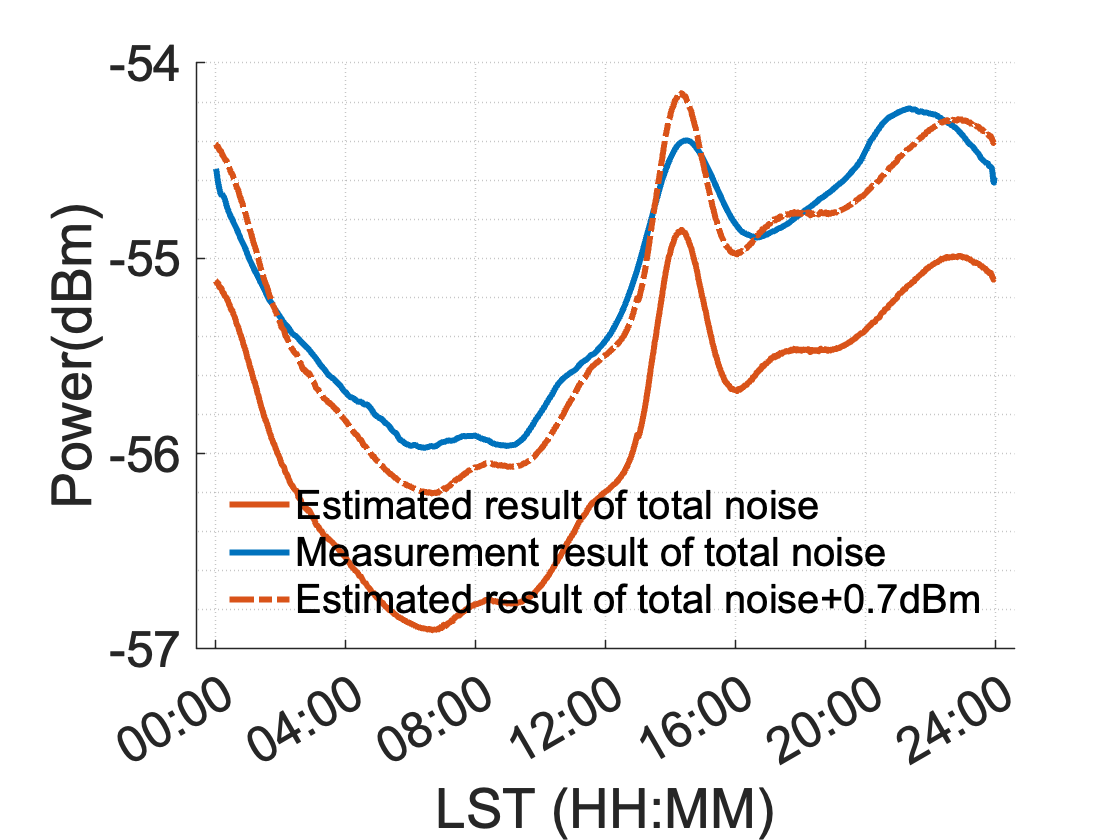}\\[1ex]
  (b)\\[2ex]
    \includegraphics[width=3.0in]{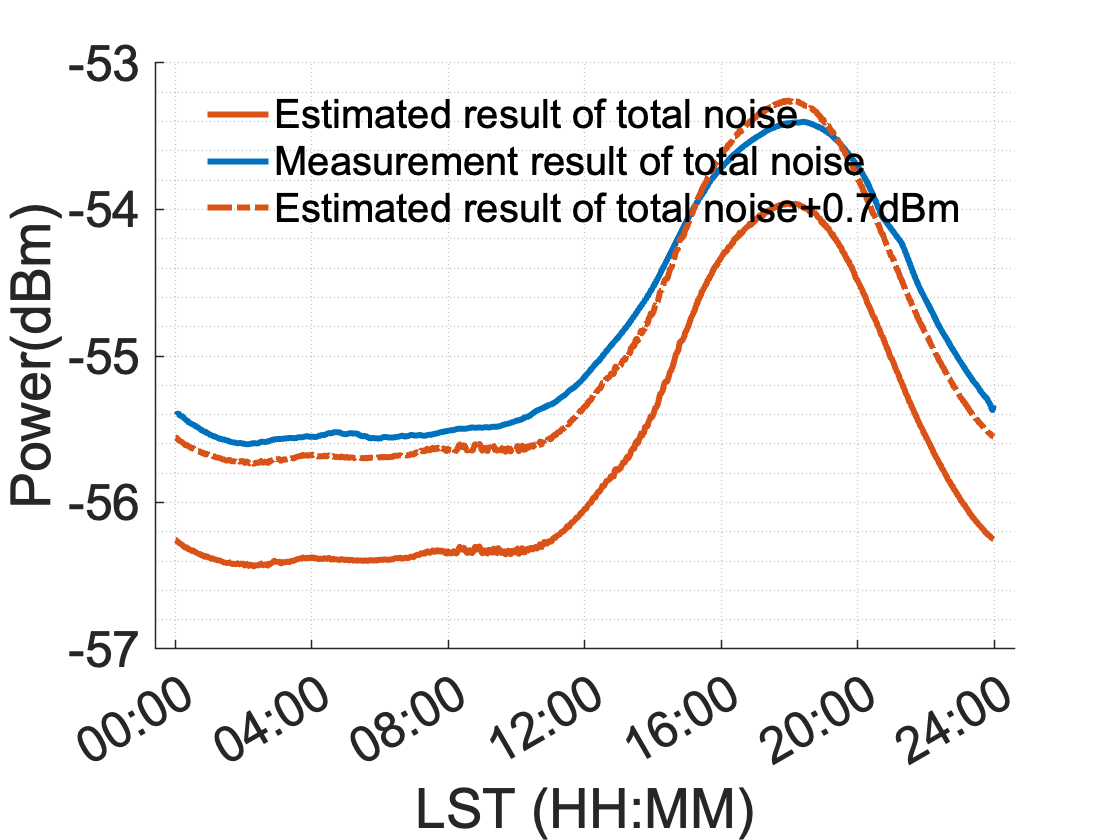}\\[1ex]
  (c)
  \caption{Simulation and test results of system noise. 
(a) Power spectral density (PSD) at the ADC input port: the yellow dotted line denotes the sky background noise, the purple dotted line denotes the expected internal system noise, the red solid line denotes the expected total noise spectrum, and the blue solid line denotes the measured noise spectrum. 
(b) Time-varying energy integral of the north--south polarization channel in the 60--90~MHz band: the blue solid line denotes the measured value, the red solid line denotes the simulated value (Galactic noise + internal noise), and the red dotted line denotes the expected value with an additional 0.7~dB residual. 
(c) Time-varying energy integral of the east--west polarization channel, with the same conventions as in (b).}
  \label{fig:system_noise}
\end{figure}

To further characterize the sky background contribution, the simulated PSD is integrated either over time or within selected frequency bands. The resulting daily averaged sky-noise spectrum is shown as the yellow dotted curve in Fig.~\ref{fig:system_noise}(a), providing a reference for the stationary noise component. In addition, integrating the PSD within the 60--90~MHz band at different local times yields the expected temporal modulation of the received power induced by Earth's rotation. This diurnal variation is shown as the blue solid curves in Fig.~\ref{fig:system_noise}(b) for the north--south polarization and in Fig.~\ref{fig:system_noise}(c) for the east--west polarization.

Fig.~\ref{fig:system_noise}(a) compares the simulated sky background noise, the expected internal system noise, and the measured noise spectrum at the ADC input. For the GRAND prototype used here as a demonstration, the peak sky-noise PSD reaches approximately $-70$~dBm during observations on 25--26 Apr. 2025 (UTC). Given the ADC full-scale range ($\pm0.9$~V) and 14-bit resolution, this corresponds to an RMS voltage of approximately 13 least significant bits, consistent with standard ADC noise characterization~\cite{Alegria2005,Blair1994}. This sky-dominated noise level defines the quantitative design benchmark for the detector: all reducible noise and interference contributions must be suppressed to values close to or below this reference.

Based on the sky-noise modeling described in Sec.~III, the simulated Galactic background power integrated over the 51--120~MHz band varies between approximately $-75$~dBm and $-70.5$~dBm, depending on local sidereal time.
Within the same band, the intrinsic electronic noise of the complete RF chain is measured to remain below $-75$~dBm.
Further reduction of internal noise would therefore yield only marginal improvements once combined in quadrature with the dominant sky noise, while significantly increasing system complexity and cost.
For this reason, a system noise power of $-75$~dBm in the core operational band is adopted as a practical and physically motivated noise-control target for the GRAND detector units.

\section{Design and Experimental Validation of Noise and Interference Suppression}

A structured four-stage optimization strategy is adopted to suppress system noise and external interference in self-triggered radio detector units. The approach consists of: 
(i) RF-chain and front-end board (FEB) low-noise prototyping, where the intrinsic RF-chain noise is minimized and characterized using controlled indoor measurements; 
(ii) suppression of non-RF-chain self-generated noise, in which the low-noise RF system developed in step (i) is employed to identify and mitigate electromagnetic contamination originating from auxiliary subsystems; 
(iii) continuous-wave (CW) interference mitigation, where on-site measurements guide the optimization of analog and digital filtering strategies to suppress persistent external emissions; and 
(iv) transient pulse-interference rejection, which addresses non-stationary external sources through data-driven classification and selection methods. 
This section details the first three stages, while the treatment of transient pulse interference is addressed separately in Sec.~VI.

\subsection{RF-Chain and Front-End Board Design for Low-Noise Operation}

\subsubsection{Analytical RF-Chain Noise Budget}

In practical RF systems, intrinsic electronic noise degrades the achievable sensitivity and must be carefully budgeted. In this work, the RF-chain noise is quantified using an equivalent noise-temperature formalism referenced to the ADC input, which enables direct comparison with the sky-noise benchmark established in Sec.~III. The analytical results are subsequently validated through dedicated laboratory measurements.

\begin{multline}
T_{RF\_chain} = T_{\text{Pre-LNA}} G_{LNA} + T_{LNA} 
+ \frac{T_{cable}}{G_{LNA}} 
+ \frac{T_{VGA}}{G_{LNA}/L_{cable}} \\
+ \frac{T_{filter}}{G_{LNA} G_{VGA}/L_{cable}} 
+ \frac{T_{Balun}}{(G_{LNA} G_{VGA})/(L_{cable} L_{filter})}.
\label{eq:Trflink}
\end{multline}
where $T$ denotes the component noise temperature, $G_{LNA}$ is the gain of the LNA (the only active device), and $L$ is the loss factor of the passive device. The front-end noise temperature described by Eq.~\eqref{eq:Trflink} is usually dominated by the LNA. The noise contribution at the ADC level of component $i$ is:

\begin{equation}
V^{i}_{ADC} = 
\sqrt{\left[ k T_i B \times 
\prod_{j=i+1}^{7} G_j \right] Z_{load}},
\label{eq:ViADC}
\end{equation}

Here, $B$ represents the bandwidth, and $G$ denotes the gain (or loss for passive devices). 
As in Sec.~III-A, the indices $i=1\dots 7$ correspond to Balun1, matching network, LNA, cable, VGA, filter, and Balun2, respectively. 
Eq.~\eqref{eq:ViADC} shows that placing low-noise components as early as possible in the electronic chain and reducing their noise factors are the two primary drivers for lowering the overall noise level of the RF chain.

\subsubsection{Low-noise amplifier design}

\begin{figure}[!t]
  \centering
  \includegraphics[width=3.0in]{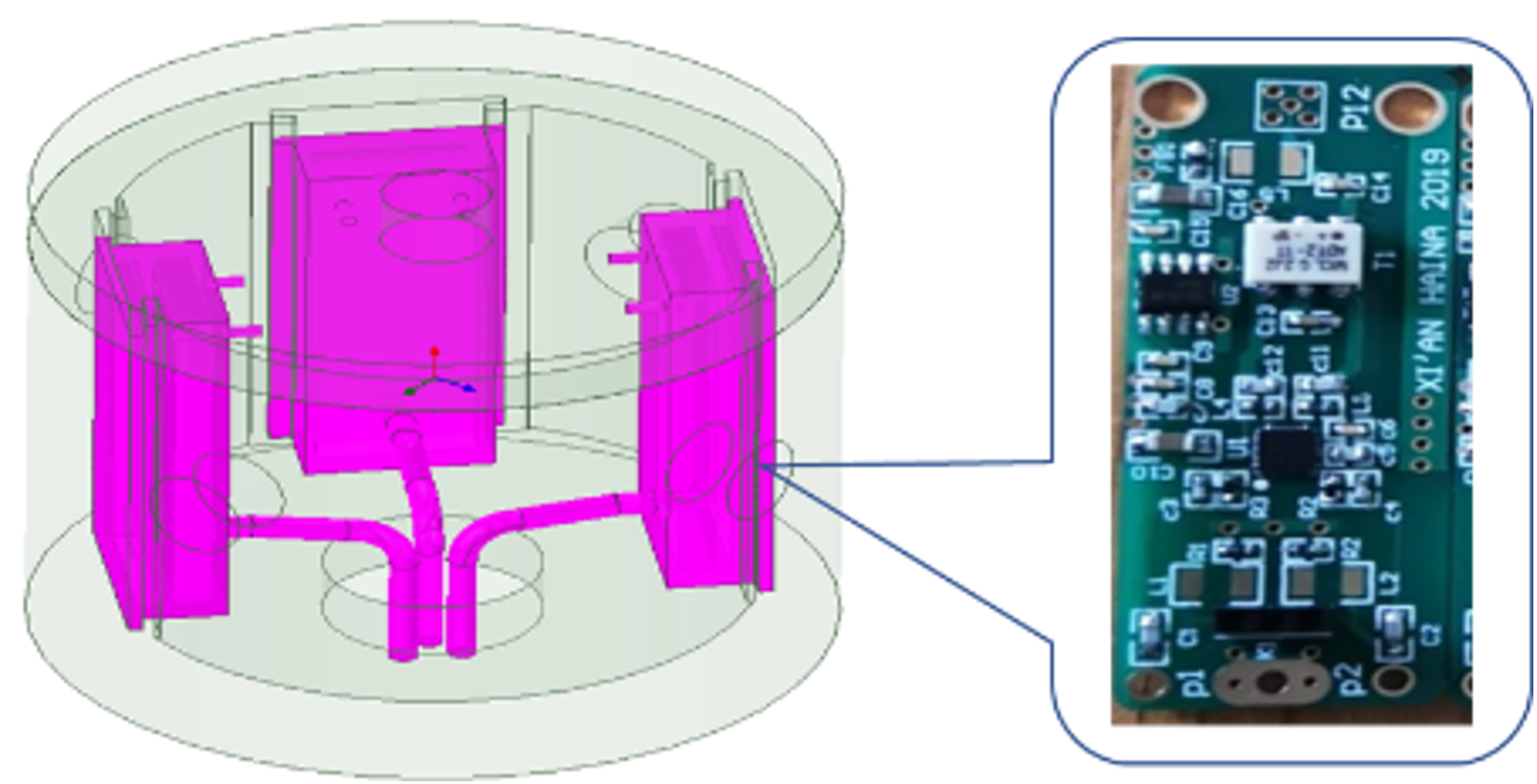}\\[1ex]
  (a)\\[2ex]
  \includegraphics[width=3.0in]{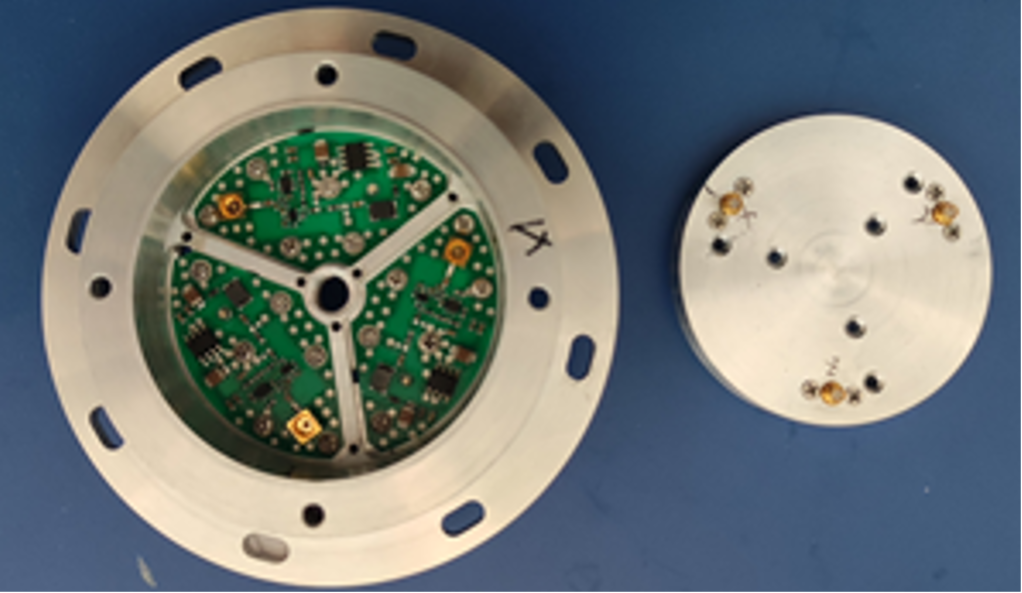}\\[1ex]
  (b)\\[2ex]
  \includegraphics[width=3.0in]{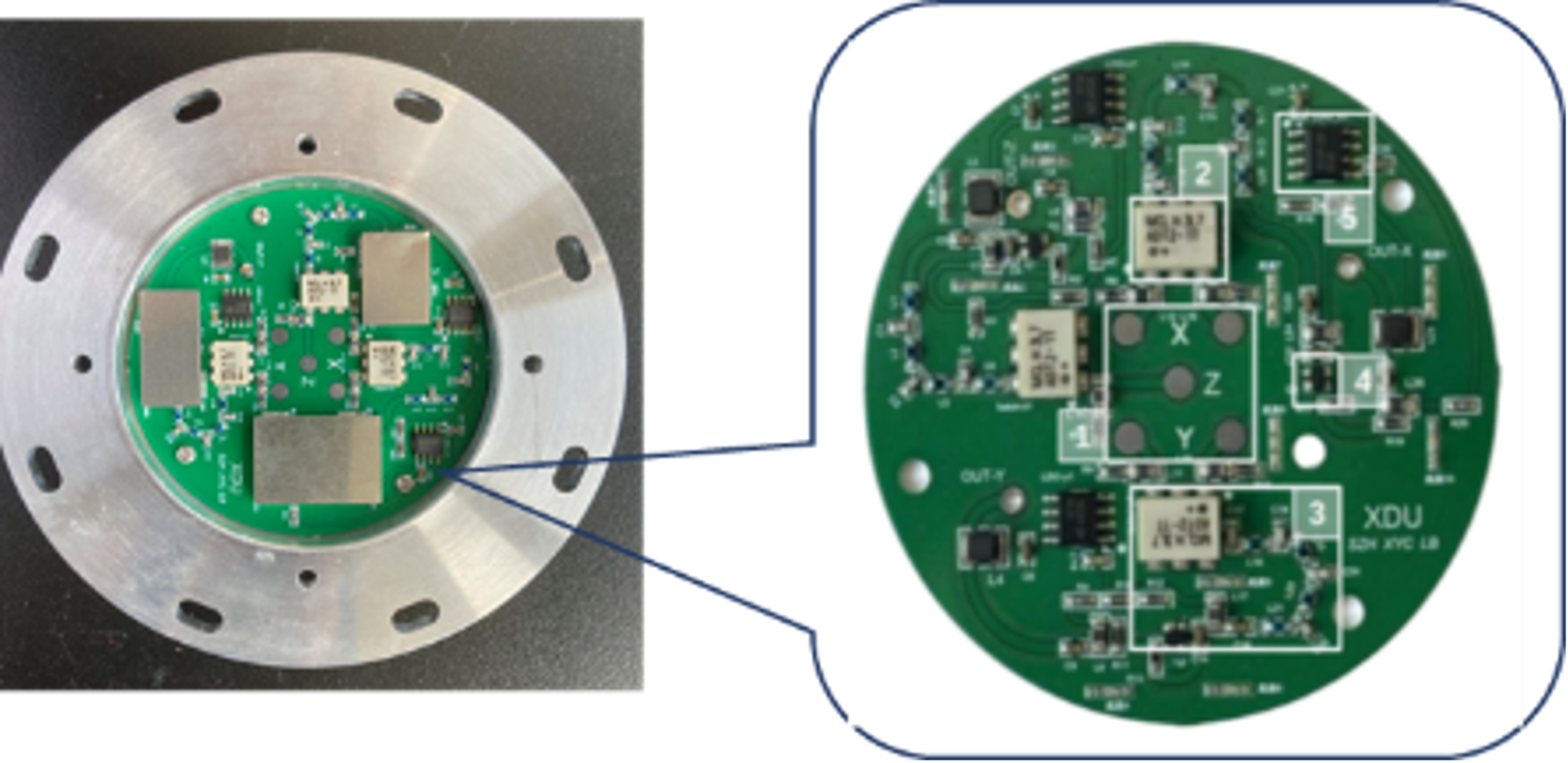}\\[1ex]
  (c)
  \caption{LNA packaging and shielding strategies developed within the GRAND framework:
  (a) differential amplifier with separated cavity;
  (b) single-ended amplifier with multi-layer interconnection and separated cavities;
  (c) single-layer single-ended amplifier with integrated EMC shielding.}
  \label{fig:LNA}
\end{figure}

Fig.~\ref{fig:LNA} compares three LNA packaging and shielding strategies developed within the GRAND framework. These designs illustrate practical trade-offs between channel isolation, electromagnetic compatibility, manufacturing complexity, and achievable noise performance.

Method~(a) is based on a differential amplifier with a separated cavity;
method~(b) employs a single-ended amplifier with multi-layer interconnection and separated cavities for the three channels;
and method~(c) adopts a single-ended amplifier with a compact layout and integrated electromagnetic compatibility (EMC) shielding for the three channels.
Considering both cost and noise performance, method~(c) was selected for the GRAND prototype.
Testbench measurements indicate that this design maintains a noise figure below 1~dB over most of the operational band (50--220~MHz), with a maximum value of approximately 1.3~dB.

In this implementation, the LNA enclosure does not form a fully enclosed metallic shielding box, since the upper part of the structure is realized using a nylon block that provides mechanical support for the antenna arms and facilitates cable routing. To mitigate electromagnetic radiation generated by the LNA and its potential coupling back into the antenna, a localized electromagnetic compatibility (EMC) shielding structure is implemented directly on the PCB around the LNA circuitry, as shown in Fig.~\ref{fig:LNA}(c). This localized shielding is designed to reduce near-field radiation from the amplifier and to limit the coupling of LNA-generated noise into the antenna, contributing to stable and reproducible noise performance at the system level.

\paragraph{Schematic and design principle}

\begin{figure*}[ht]
  \centering
  \includegraphics[width=7.16in]{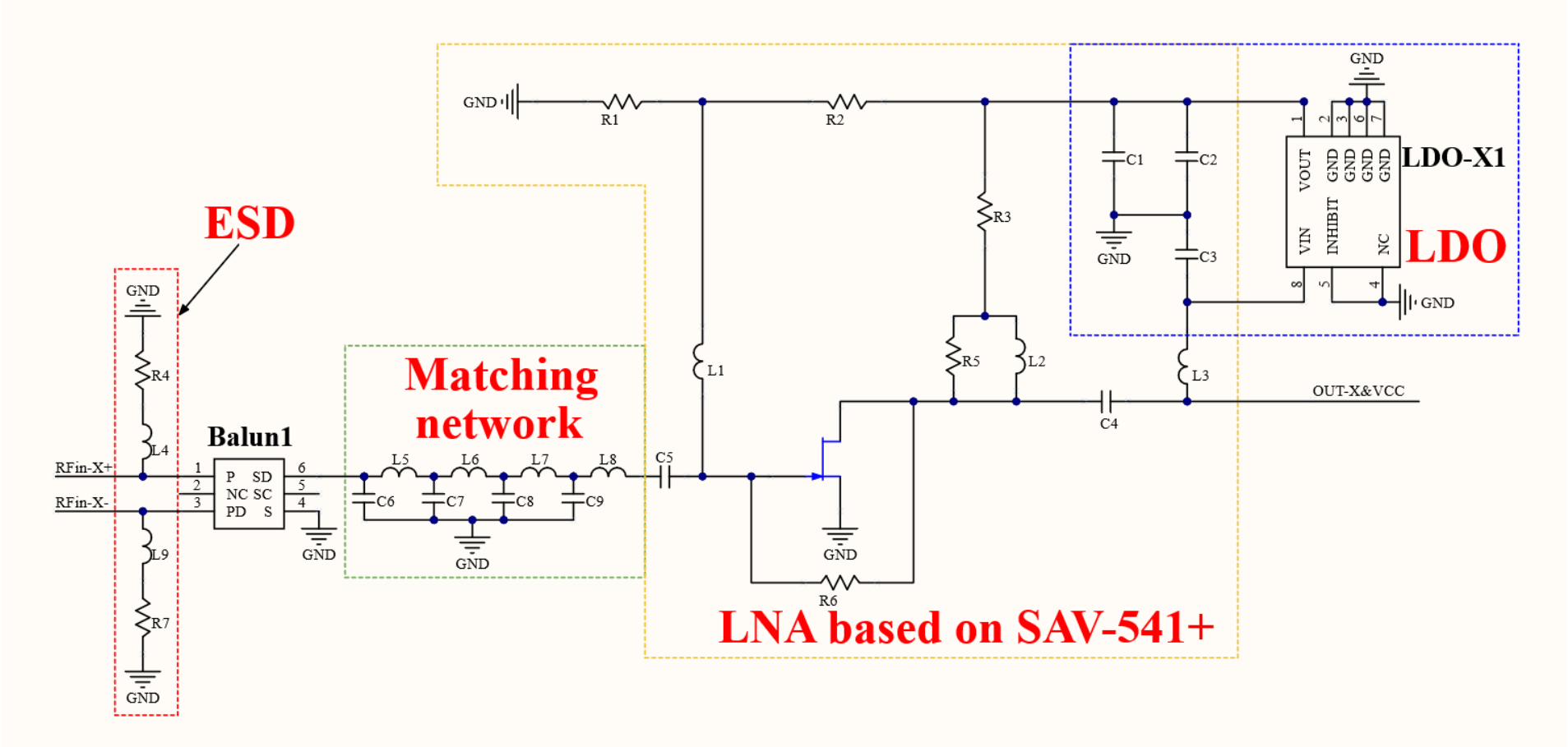}
  \caption{Schematic of the designed low-noise amplifier (LNA), including the input/output matching networks, DC bias network, feedback network, and ESD protection.}
  \label{fig:LNA_schematic}
\end{figure*}

Fig.~\ref{fig:LNA_schematic} shows the schematic of the designed LNA.
The circuit is based on a field-effect transistor as the active device and consists of an input matching network, an output matching network, a DC bias network, a feedback network, and an electrostatic-discharge (ESD) protection structure.
The LNA is powered via a Bias-T at the output port.
When supplied with a DC voltage of 5~V, each channel draws approximately 47~mA, corresponding to a power consumption of about 235~mW per channel.

The input matching network, located at the front end of the LNA, is optimized for broadband impedance matching between the antenna and the amplifier, transforming the antenna impedance to 50~$\Omega$.
This approach reduces sensitivity to impedance mismatch and simplifies the subsequent RF-chain design.
A balun is used to convert the differential antenna signal into a single-ended signal while simultaneously performing impedance transformation.

A resistive--capacitive feedback network, composed of components R6 and C10, is introduced to improve the input return loss ($S_{11}$) and to shift the minimum-noise impedance point closer to 50~$\Omega$.
In addition to improving matching, this feedback topology ensures unconditional stability across the required frequency band and enhances passband flatness.
As a result, a noise figure as low as 0.8~dB can be achieved while maintaining a stable gain of approximately 20~dB over the 50--200~MHz band.
This design represents a practical compromise between noise performance, bandwidth, gain, stability, and flatness, tailored to the requirements of self-triggered radio detection in a galactic-noise-dominated environment.

Although the overall circuit topology follows a conventional broadband low-noise design philosophy, the implementation is specifically optimized for large-scale, self-triggered radio arrays.
The same structure can be adapted to other junction bipolar transistor or field-effect transistor devices by adjusting the DC bias network according to the corresponding datasheets to reach the optimal operating point.

The noise contribution of the LNA is computed using Eq.~\eqref{eq:ViADC} with typical in-band parameters:
LNA gain = 20~dB, noise figure = 1.3~dB;
RF cable loss = 0.25~dB;
VGA gain = 20~dB, noise figure = 10~dB;
filter loss = 0.25~dB;
Balun2 primary-to-secondary coil ratio = 1:2;
ADC load impedance = 200~$\Omega$.
At a temperature of 35$^\circ$C, integration over a 190~MHz bandwidth yields an ADC-referred noise voltage of $7.05\times10^{-4}$~V, corresponding to 6.47~LSB for the selected ADC, where $\text{LSB}=1.8~\text{V}/2^{14}\approx110~\mu$V.
Within the frequency band of interest, this thermal noise contribution remains below the dominant Galactic background noise, satisfying the sensitivity requirements.
The resulting noise contribution of the \emph{nut} across the full frequency band is shown by the blue curve in Fig.~\ref{fig:packaging test result}.

Here, \emph{nut} refers to the front-end assembly extending from the antenna interface to the LNA input, including the nylon support structure fixing the antenna and, within a flange-mounted cavity, the balun, matching network, LNA, and its auxiliary circuits, and terminated by an N-type connector for cable connection. The term \emph{nut} is used as a label for this mechanically integrated front-end assembly, rather than as an acronym.

The FEB design is not the primary focus of this work; however, since it constitutes a key part of the RF chain, its main design considerations are summarized briefly.
Standardized ground filtering is applied across all circuit grounds, and EMC shielding enclosures are implemented within the instrument cavity, following the same principles as the LNA shielding shown in Fig.~\ref{fig:LNA}(c).
The GRANDProto FEB integrates a VGA with an adjustable gain range from $-10$ to $+20$~dB (typically operated at maximum gain) and an analog filter implementing a 30--230~MHz passband.
A four-channel ADC digitizes the signals with 14-bit resolution over a $\pm0.9$~V range at 500~MSample/s.
Testbench measurements indicate an RMS FEB-internal noise contribution of approximately 4~LSB at the ADC input for each channel.
As shown in Fig.~\ref{fig:packaging test result}, this contribution remains well below the LNA-referred noise level over the effective 45--220~MHz band, making the FEB suitable for system-level noise validation and deployment in the GRANDProto project.
\subsubsection{Experimental Validation of the RF-Chain Noise Budget}

To experimentally validate the RF-chain noise model, dedicated measurements of the internal noise contribution were performed with the antenna removed. The \emph{nut} configuration includes all components from the antenna interface to the LNA output, ensuring that the measured spectrum reflects intrinsic RF-chain behavior rather than environmental pickup. This differential measurement protocol provides a reproducible and end-to-end method to quantify the effective noise contribution of individual RF-chain elements once integrated into the complete system, a capability that is typically inaccessible through standalone component characterization.
By coherently combining sky-background modeling, RF-chain noise budgeting, electromagnetic compatibility (EMC) design, and measurement-driven validation, this work provides, to the best of our knowledge, the first comprehensive and quantitative detector-unit-level review of noise sources and RF-chain architecture for self-triggered extensive air-shower radio experiments.

\begin{figure}[!t]
  \centering
  \includegraphics[width=3.0in]{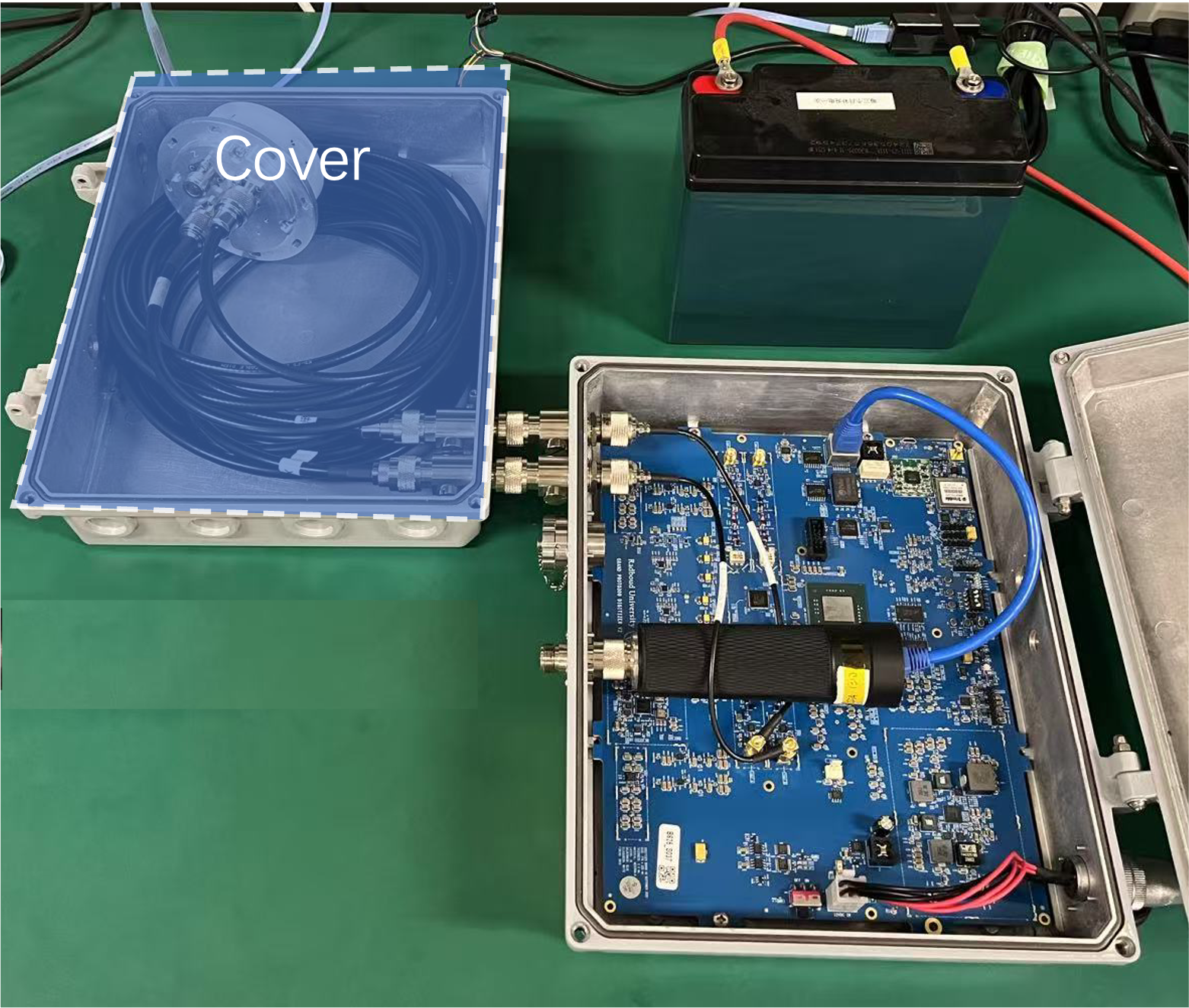}
  \caption{Schematic diagram of metal box packaging test}
  \label{fig:packaging test}
\end{figure}

\begin{figure}[!t]
  \centering
  \includegraphics[width=3.0in]{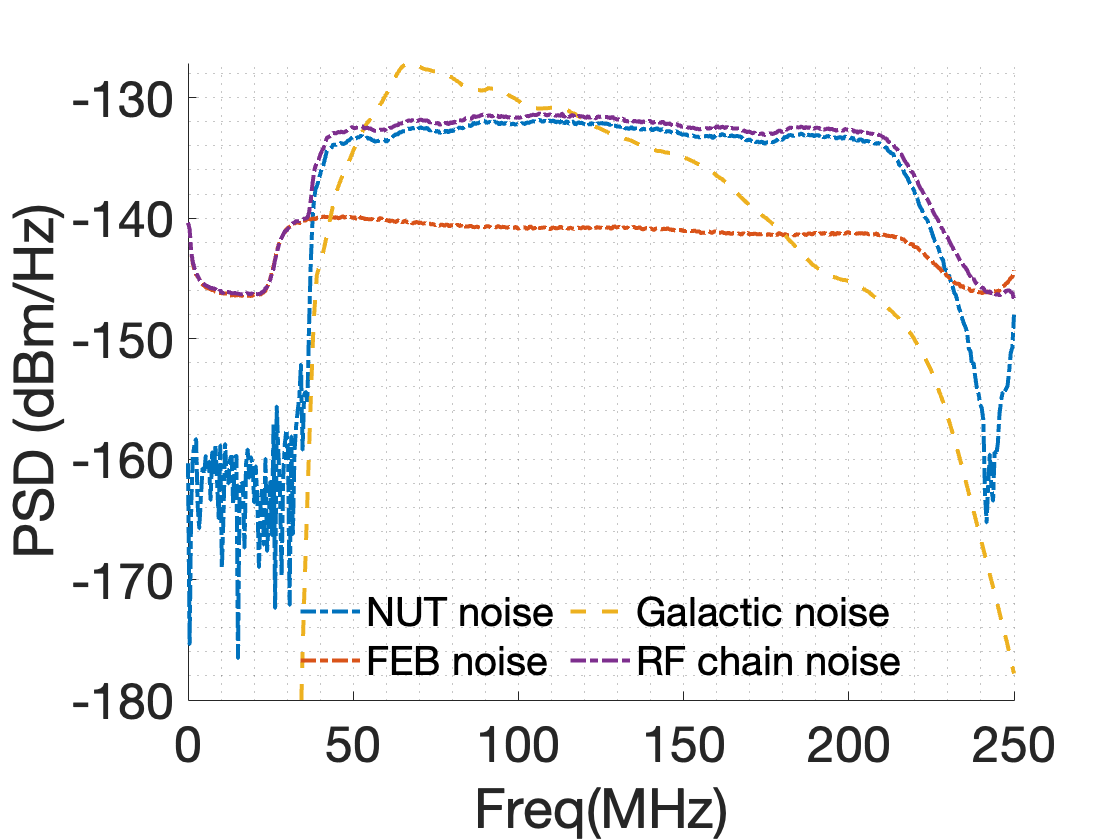}
  \caption{Test results of noise contribution of RF chain (\emph{nut} + FEB) (Galactic background noise is provided for comparison)}
  \label{fig:packaging test result}
\end{figure}

Caution is required: merely removing the antenna arm leaves the fixed flange susceptible to ambient noise coupling through residual grounding paths. For accurate measurement, all non-essential components should be removed and the RF-chain assembly placed within a dedicated metal shielding enclosure during the test, as shown in Fig.~\ref{fig:packaging test}. The complete RF-chain noise spectrum is shown as the blue curve in Fig.~\ref{fig:packaging test result}. A comparison with the simulated sky background noise (purple dashed line), computed in Sec.~III-C, shows that the internal noise level remains below the astronomical background within the 50--120~MHz core observational band, thereby achieving the design target of detector noise approaching the fundamental Galactic noise floor.

A direct standalone measurement of the LNA noise temperature is impractical once the LNA is integrated into the full RF chain.
Instead, an effective LNA noise contribution is inferred indirectly by comparing two internal-noise measurements performed at the ADC level:
(i) the internal noise of the complete RF chain including the LNA, and
(ii) the internal noise of the FEB alone with the LNA excluded.
By subtracting the FEB-only contribution in quadrature, the effective LNA-related noise contribution at the ADC input is obtained.

The length of the coaxial cable connecting the LNA and the FEB is dictated by the mechanical layout of the detector unit.
In the GRAND prototype, the LNA is mounted at the top of a $\sim$3~m mast together with the antenna, while the FEB is installed beneath the solar panel.
Using semi-rigid coaxial cables would significantly complicate installation, maintenance, and large-scale field deployment.
The selected flexible coaxial cable introduces less than 0.5~dB attenuation over the full 30--250~MHz band, which has a negligible impact on the overall system noise budget.

The FEB enclosure used in the prototype system is made of cast aluminum rather than plastic.
This metallic enclosure provides effective electromagnetic shielding.
To further improve shielding effectiveness, conductive gaskets are installed at the interface between the enclosure body and the lid, ensuring electrical continuity.
All connectors are electrically bonded to the enclosure, such that the FEB functions as a fully shielded unit against external and self-generated electromagnetic interference.

\subsection{Suppression of Self-Generated Non-RF-Chain Noise}

\subsubsection{Selection and EMC Evaluation of Core Functional Devices}

Although many commercial components satisfy the functional requirements of a self-triggered detector unit, their electromagnetic emission characteristics are rarely specified in datasheets and can critically affect low-frequency radio measurements. We therefore performed systematic screening tests of candidate devices using the setup described in Sec.~4.1, focusing on both transient-pulse and continuous-wave contamination.

\begin{figure*}[!t]
\begingroup
  \captionsetup[subfloat]{labelformat=empty} 
  \centering
  \newcommand{\fivecolwidth}{\dimexpr\linewidth/5-0.8em\relax} 
  \subfloat[a1]{\includegraphics[width=\fivecolwidth,height=1.25in]{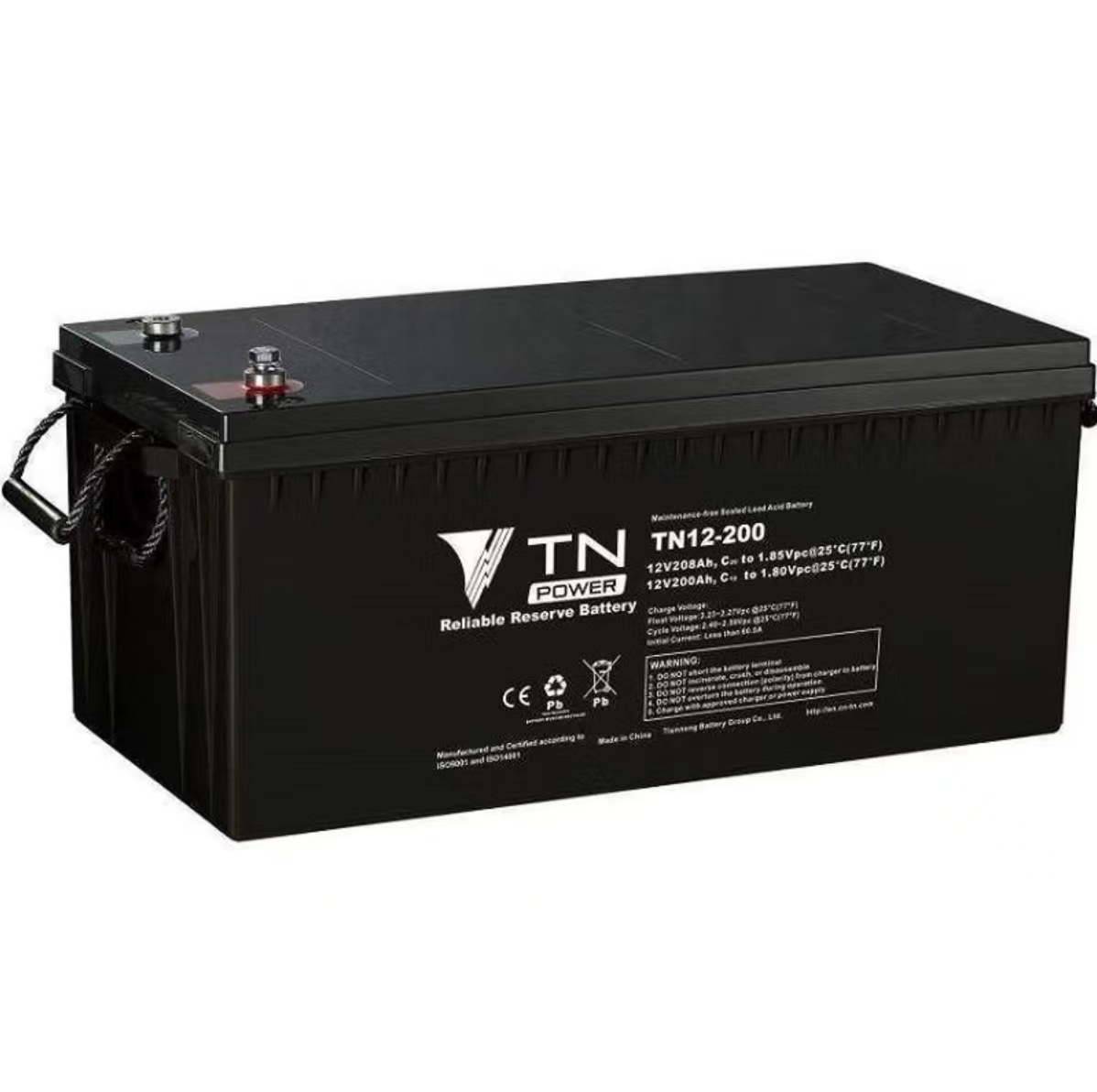}}\hfill
  \subfloat[b1]{\includegraphics[width=\fivecolwidth]{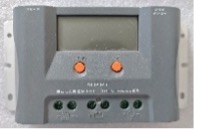}}\hfill
  \subfloat[c1]{\includegraphics[width=\fivecolwidth,height=1.25in]{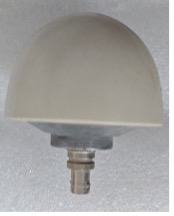}}\hfill
  \subfloat[d1]{\includegraphics[width=\fivecolwidth,height=1.25in]{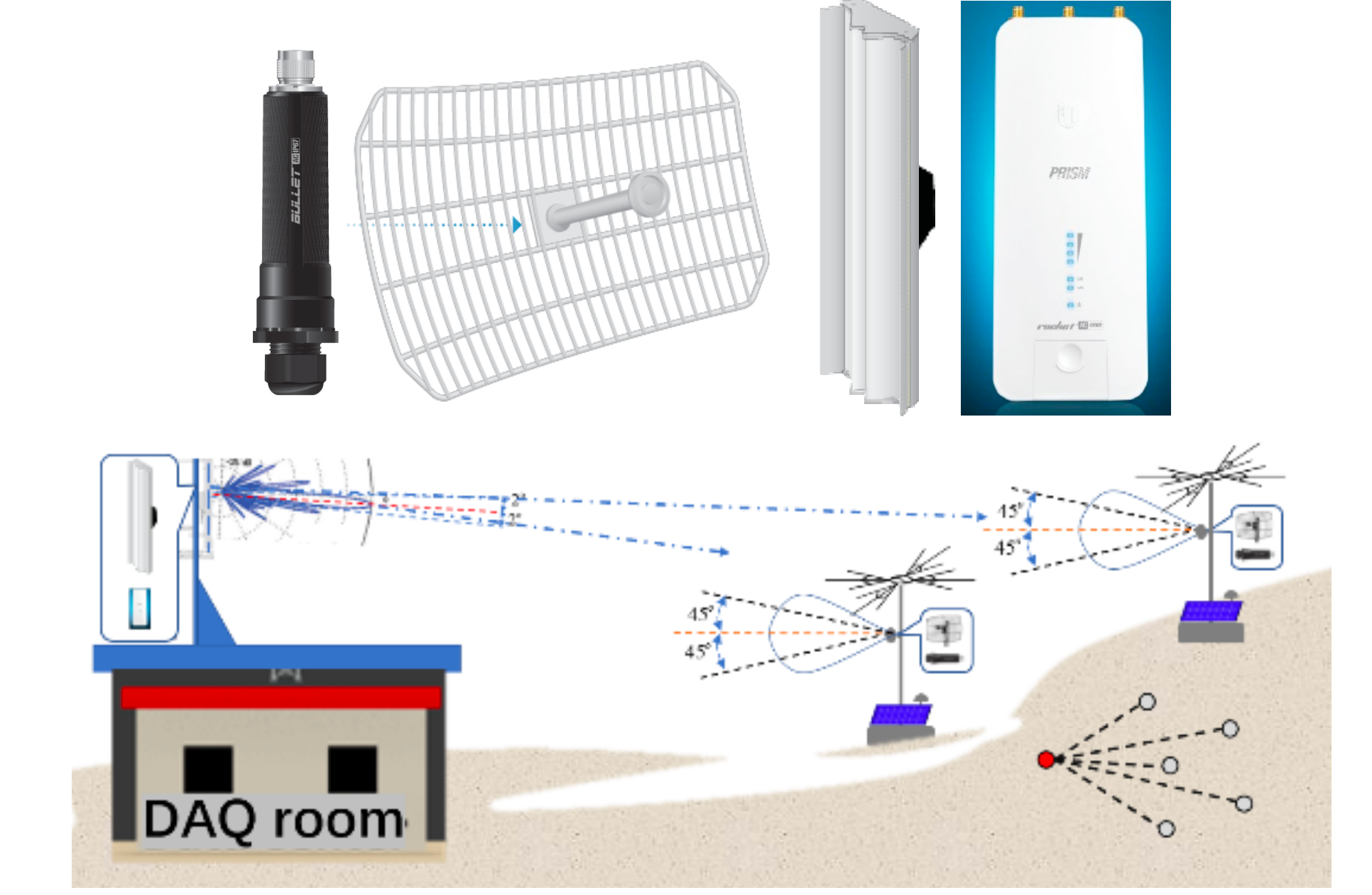}}\hfill
  \subfloat[e1]{\includegraphics[width=\fivecolwidth,height=1.25in]{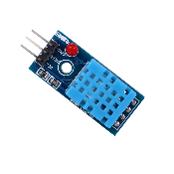}}\\[0.8ex]

  \subfloat[a2]{\includegraphics[width=\fivecolwidth]{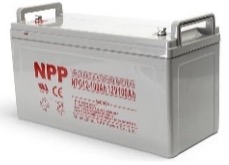}}\hfill
  \subfloat[b2]{\includegraphics[width=\fivecolwidth]{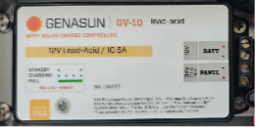}}\hfill
  \subfloat[c2]{\includegraphics[width=\fivecolwidth,height=1.25in]{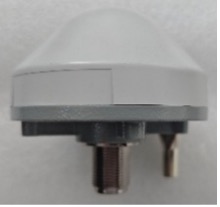}}\hfill
  \subfloat[d2]{\includegraphics[width=\fivecolwidth]{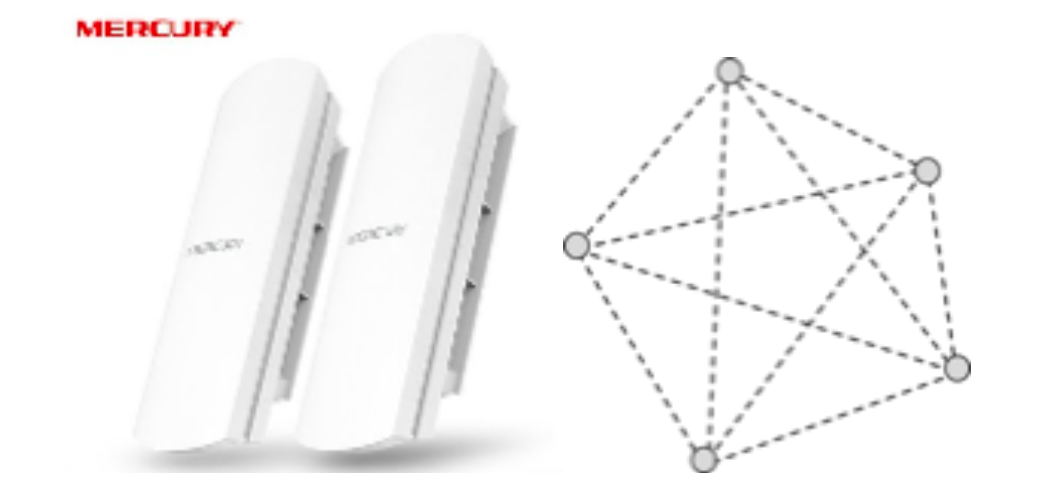}}\hfill
  \subfloat[e2]{\includegraphics[width=\fivecolwidth]{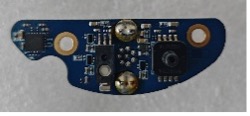}}

  \caption{Commercial devices for core functional modules in detector units:
  (a1)--(a2) batteries; (b1)--(b2) charge controllers; (c1)--(c2) GPS antennas;
  (d1)--(d2) communication systems; (e1)--(e2) slow-control sensors.}
  \label{fig:commercial_devices}
\endgroup
\end{figure*}

Datasheets rarely provide specifications related to radio-frequency interference (RFI), as such characteristics are seldom a primary concern for most applications. Consequently, we experimentally evaluated several commercial products using the setup described in the previous section. For battery selection (Fig.~\ref{fig:commercial_devices}(a)), RFI is not a limiting factor; instead, capacity as well as high- and low-temperature performance are the dominant criteria for reliable outdoor operation at the GRANDProto300 site in the Gobi Desert.  

Charge controllers frequently employ pulse-management techniques to improve charging efficiency, which can generate substantial short-duration pulses. In addition, their integrated monitoring systems and LCD displays (Fig.~\ref{fig:commercial_devices}(b1)) are found to introduce significant RFI. Although Fig.~\ref{fig:commercial_devices}(b2) shows a charge controller designed with dedicated RFI suppression, our measurements indicate that residual transient pulses persist, indicating the need for further optimization.  

The GPS antenna, together with its associated receiver chip, provides position information and pulse-per-second timing for trigger timestamping. As shown in Fig.~\ref{fig:commercial_devices}(c1), the two tested antenna types exhibit negligible performance differences under current experimental conditions. In contrast, the design of the communication subsystem is considerably more critical. A master--slave network architecture (Fig.~\ref{fig:commercial_devices}(d1)) introduces unequal latency between near and distant detection units, whereas a distributed multipoint-to-multipoint network (Fig.~\ref{fig:commercial_devices}(d2)) requires omnidirectional duplex antennas at each station, increasing power consumption and introducing relay-induced delays. Both architectures affect rapid self-triggering performance; however, by employing high-gain antennas, the master--slave configuration is favored for GRANDProto300, achieving communication ranges of up to 20~km.  

Nevertheless, experimental tests confirm that wireless communication inevitably introduces strong electromagnetic noise, necessitating dedicated suppression measures. For slow-control sensors, the power-consumption difference between analog and digital options is marginal, but significant noise originating from internal power-management chips and clock circuits is observed. Therefore, unless strict local monitoring is required, such sensors should be removed from individual detection units and replaced by a limited number of standalone sensors for centralized environmental monitoring.


\subsubsection{System-Level EMC Test and Noise Coupling Analysis}

\begin{figure}[!t]
  \centering
  \includegraphics[width=3.0in]{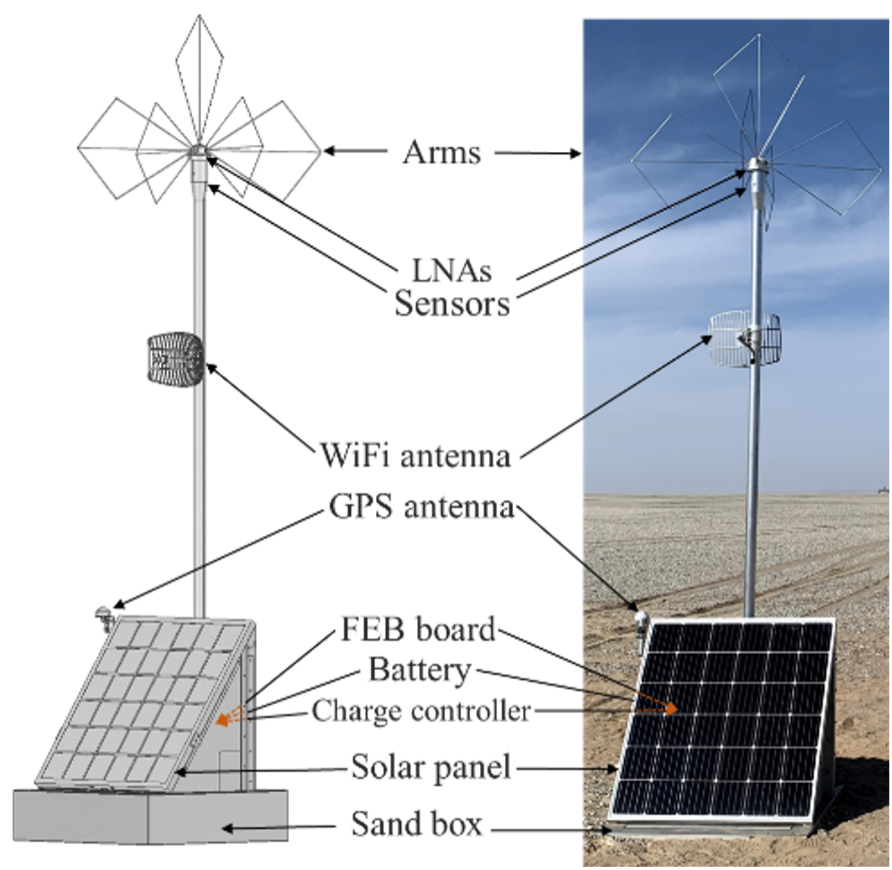}
  \caption{Detection units used in the prototype tests (see also the general overview in \cite{clery2025}).}
  \label{fig:science}
\end{figure}

During the initial tests, we chose devices (a1), (b2), (c2), (d1), and (e2), together with a tri-polarized antenna designed by Xidian University for the GRAND prototype and the FEB, to build the test units shown in Fig.~\ref{fig:science}.

Initial field tests using the selected components revealed noise levels significantly exceeding the sky-noise benchmark established in Sec.~III, confirming the presence of strong self-generated electromagnetic contamination. This observation motivated a systematic EMC investigation aimed at identifying dominant coupling paths and quantifying their contributions.

\begin{figure}[!t]
  \centering
  \includegraphics[width=3.0in]{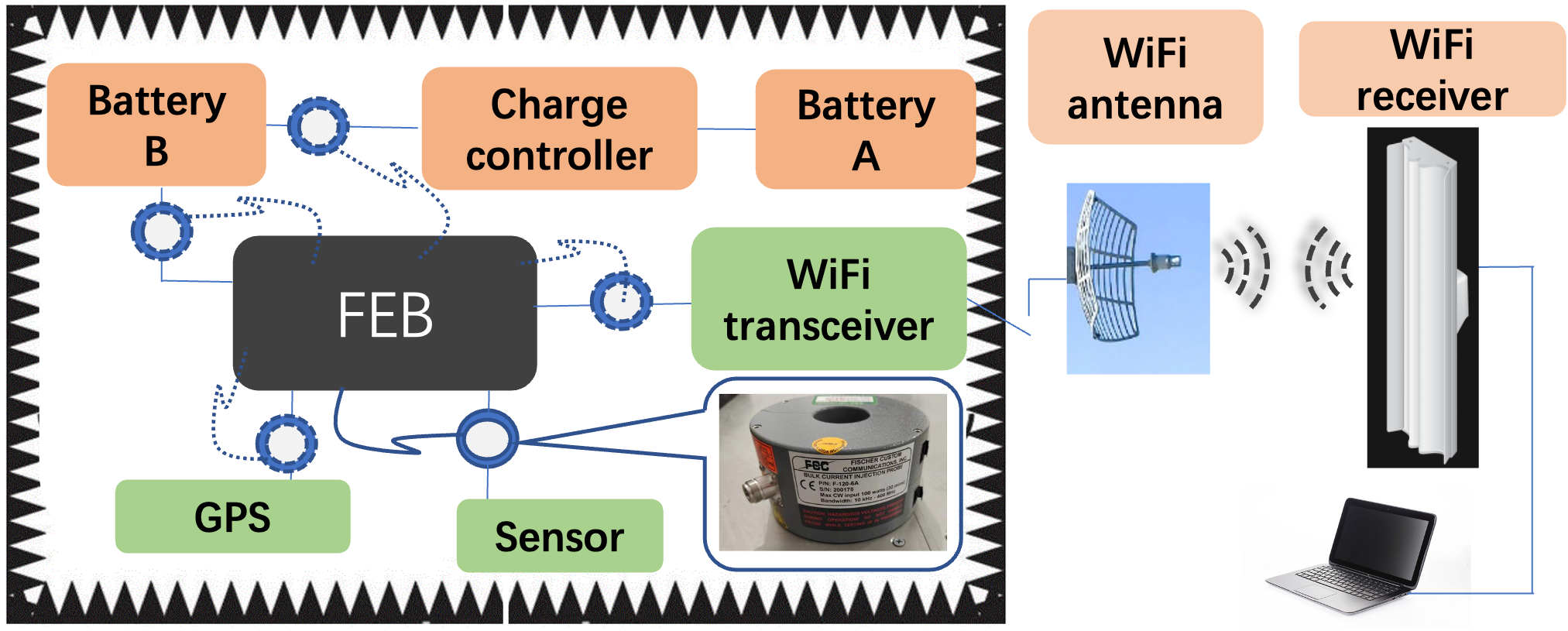}\\
  (a)\\[2ex]
  \includegraphics[width=3.0in]{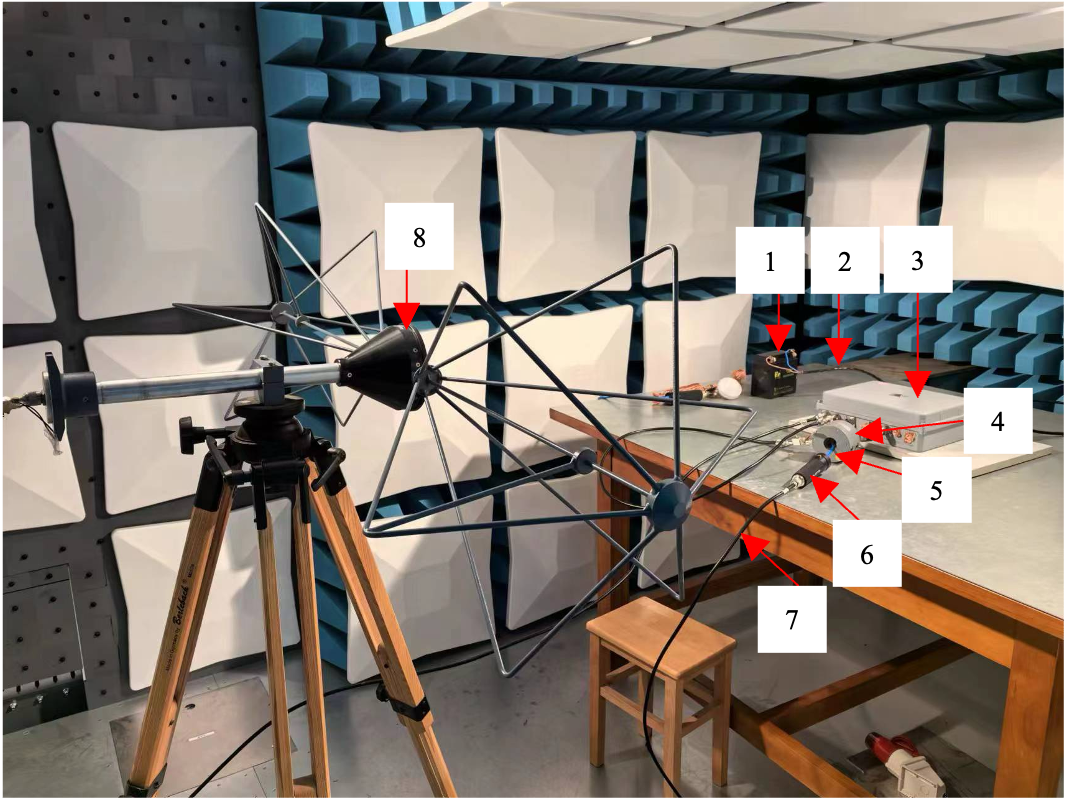}\\
  (b)\\[2ex]
  \caption{(a) Schematic diagram of the EMC shielding anechoic chamber test system (b) Physical photo of the EMC shielding anechoic chamber test}
  \label{fig:test system}
\end{figure}

As anticipated (see Fig.~\ref{fig:Schematic diagram}), cable-radiation coupling introduces substantial non-RF-chain noise into the reception system via a self-coupling loop. To quantify these contributions, the system was relocated to an EMC anechoic chamber and tested with the configuration shown in Fig.~\ref{fig:test system}, incorporating four critical modifications: (i) the antenna and LNA were removed; (ii) Wi-Fi transmission was isolated while connectivity was maintained through a through-wall connection to eliminate interference, thereby establishing a noise-free data acquisition chain; (iii) solar panels were replaced with a 24~V battery (A) charging the 12~V system battery (B); and (iv) coupling measurements were performed with one FEB RF input port alternating between an EMC biconical antenna (item~\#8 in Fig.~\ref{fig:test system}b) positioned at a fixed distance from the test cables (spatial radiation assessment) and a current probe (item~\#4 in Fig.~\ref{fig:test system}b) clamped around cables (conduction-current measurement). Analysis prioritized current-probe data due to its superior stability and sensitivity in the confined chamber environment and its direct coupling-measurement capability~\cite{Jarrix2010}.

\begin{figure}[!t]
  \centering
  \includegraphics[width=3.0in]{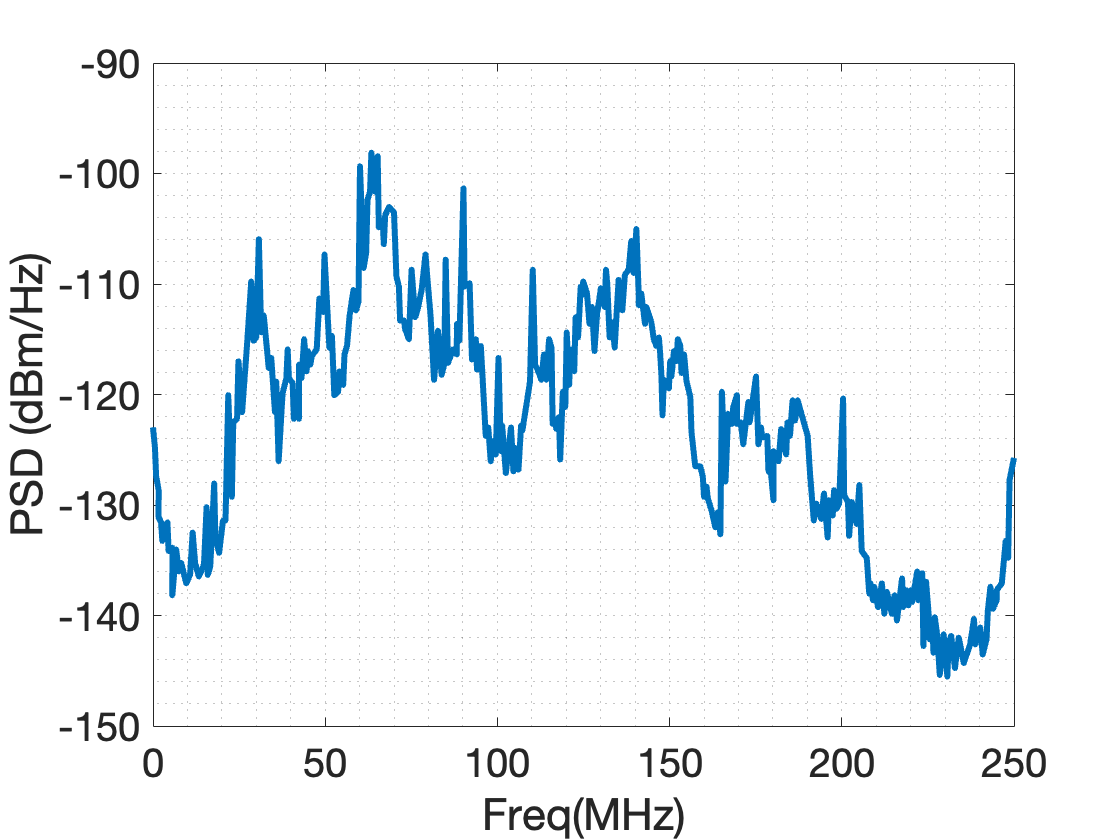}\\
  (a)\\[2ex]
  \includegraphics[width=3.0in]{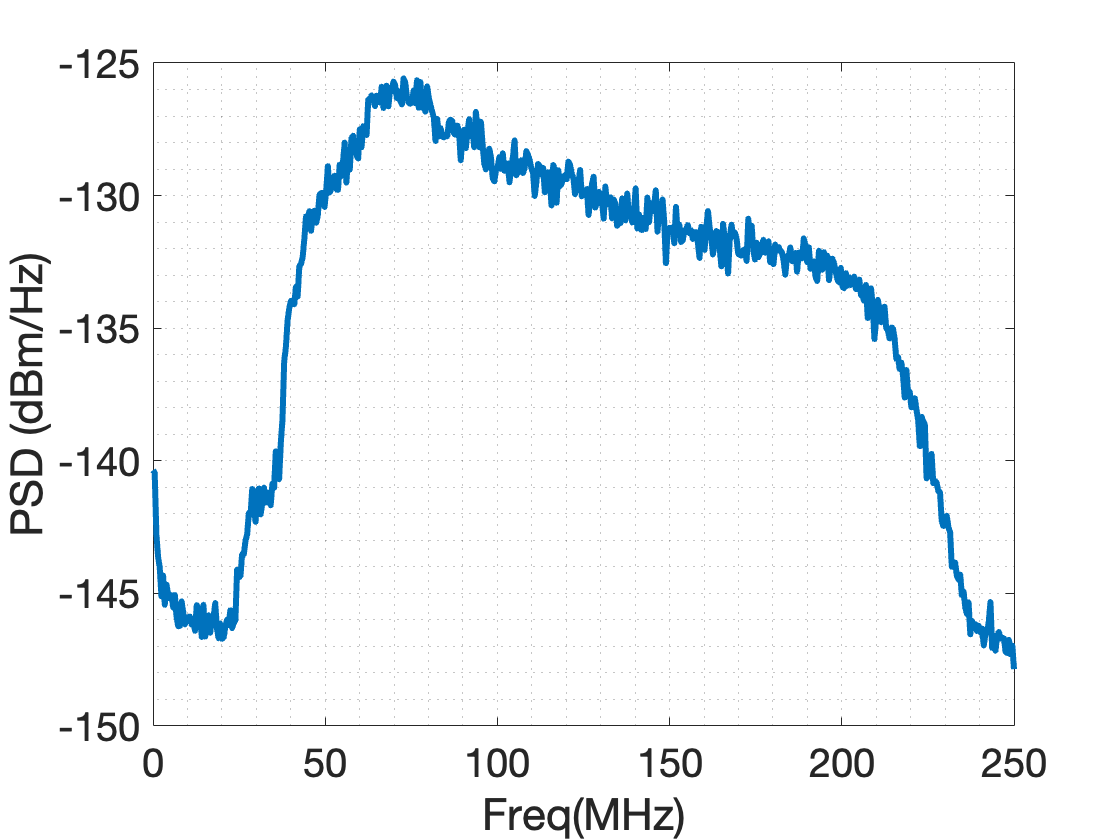}\\
  (b)\\[2ex]
  \caption{%
  System-level in-situ noise comparison of the detector unit (EW channel)
  measured at the GRANDProto300 site:
  (a) early prototype configuration prior to RF-chain and EMC optimization;
  (b) optimized detector unit after implementation of the full noise-mitigation
  strategy.}
  \label{fig:noise EMC}
\end{figure}


\begin{figure}[!t]
  \centering
  \includegraphics[width=3.4in]{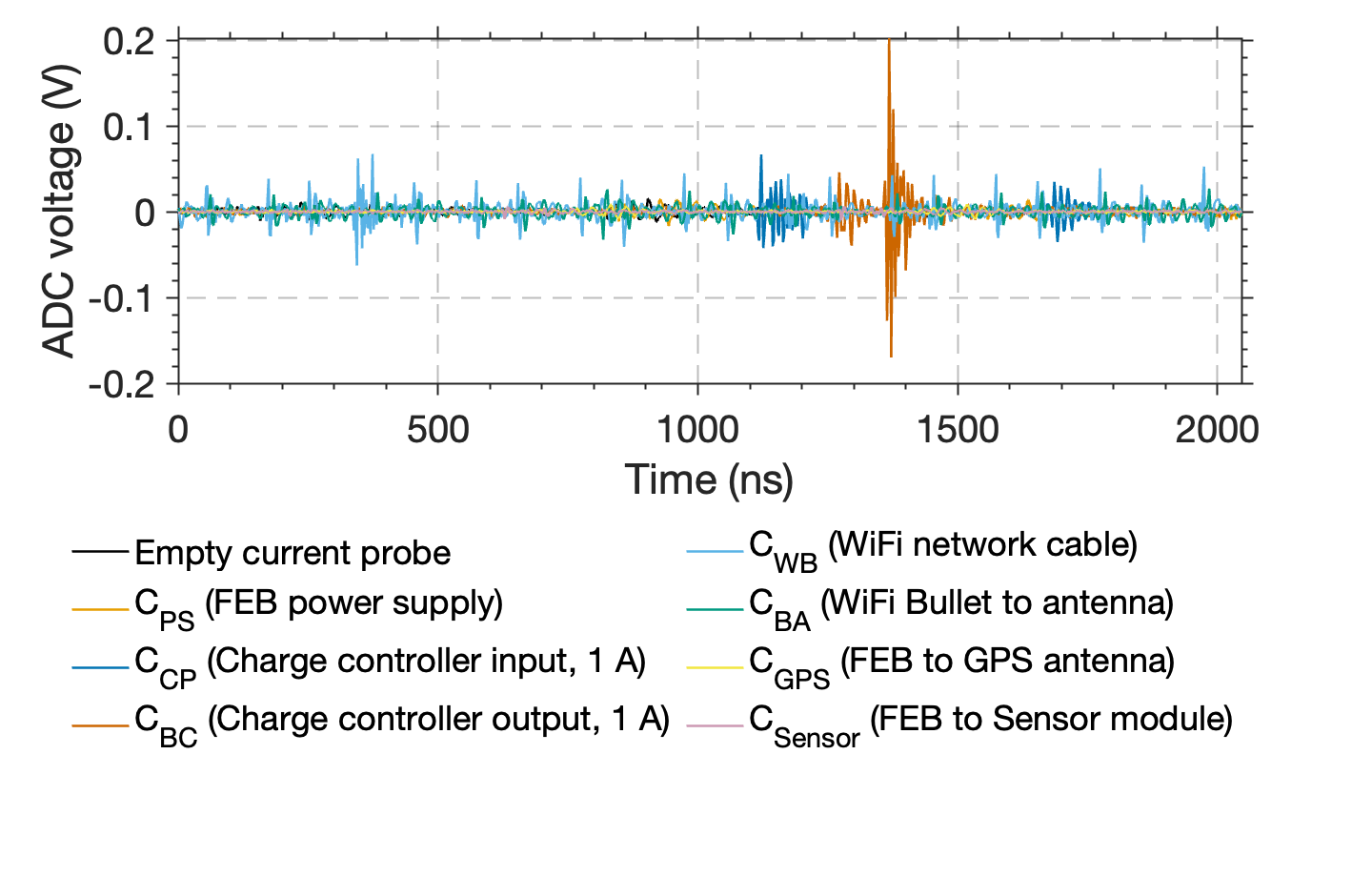}\\
  (a)\\[2ex]
  \includegraphics[width=3.4in]{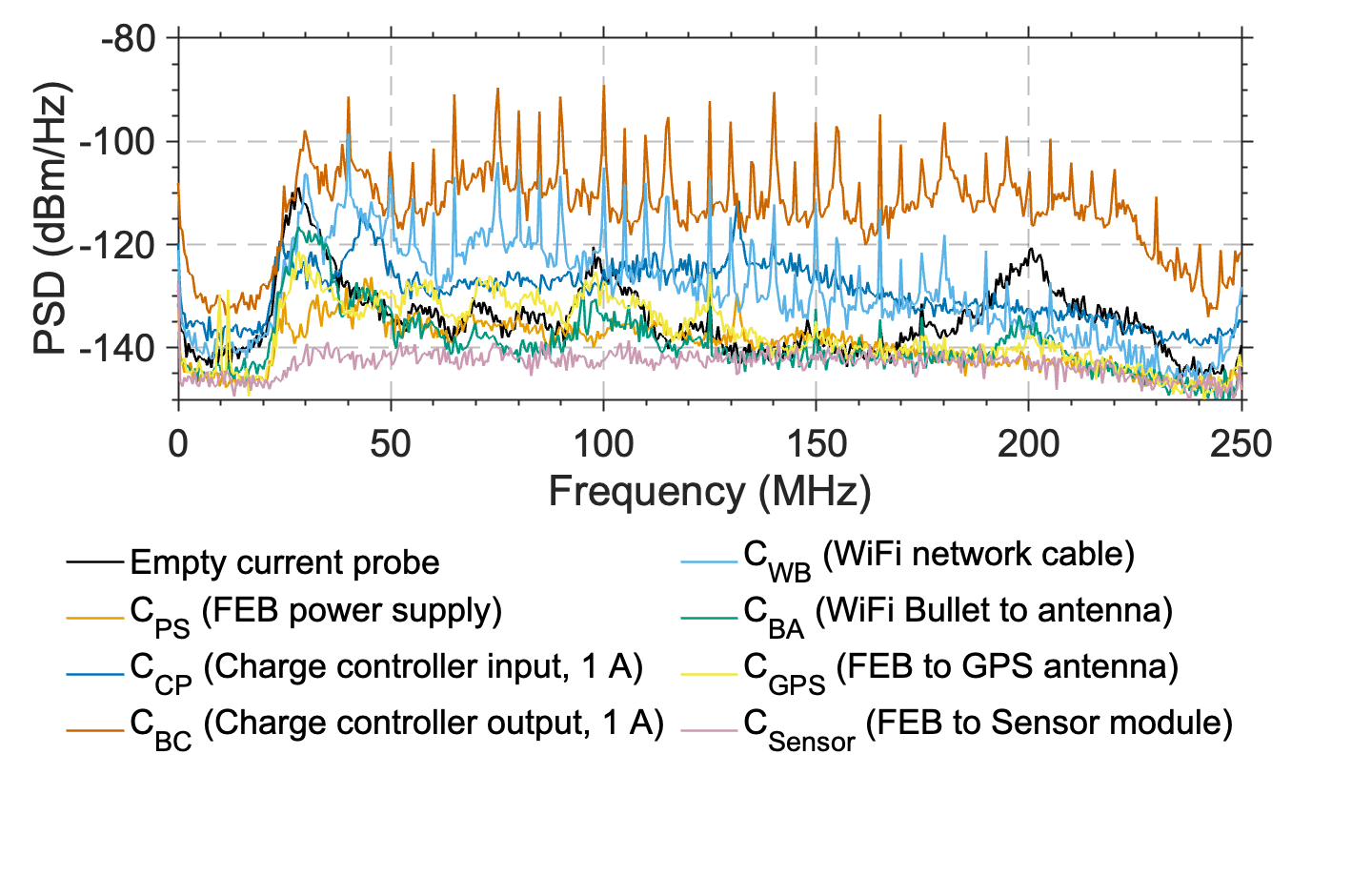}\\
  (b)\\[2ex]
  \caption{Time-domain and frequency-domain test results of RF current on different cables coupled to current probes (a) Time-domain RF current test results on different cables; (b) RF current spectrum on different cables}
  \label{fig:different cables}
\end{figure}

The current probe successively captured time-domain waveforms and frequency spectra from multiple cables, as shown in Fig.~\ref{fig:test system}(b), with the corresponding results presented in Fig.~\ref{fig:different cables}. Table~\ref{tab:noise} summarizes the different noise contributions and their time- and frequency-domain characteristics. In the time domain, the main potential sources are the charge-controller cable, the Wi-Fi network cable, and the Wi-Fi RF cable. In the frequency domain, the dominant contributors are the Wi-Fi network cable, the Wi-Fi RF cable, the FEB power-supply cable, and the charge-controller cable. Radiation from these cables must therefore be suppressed, as detailed in the next section.

\paragraph{Final in-situ system-level validation}
It should be emphasized that Fig.~\ref{fig:noise EMC} is not intended to illustrate the
effect of a single EMC mitigation step. Instead, it provides a final in-situ system-level
validation of the systematic noise-identification and mitigation protocol associated with
the anechoic-chamber measurements in Fig.~\ref{fig:different cables} and the summarized
noise characteristics in Table~\ref{tab:noise}. The reduction of the noise floor observed
in Fig.~\ref{fig:noise EMC} therefore reflects the cumulative impact of RF-chain optimization,
EMC improvements, and suppression of self-generated interference, and serves as an overall
quantitative benchmark of the progress achieved from the initial prototype to the current system.

\begin{table*}[!t]
\caption{List of noise characteristics. Acronyms are defined in the text.}
\label{tab:noise}
\centering
\renewcommand{\arraystretch}{1.25}
\renewcommand{\tabularxcolumn}[1]{m{#1}}            
\newcolumntype{Y}{>{\RaggedRight\arraybackslash}X}  
\newcolumntype{C}{>{\centering\arraybackslash}m{0.25\textwidth}} 
\begin{tabularx}{\textwidth}{|Y|Y|c|Y|C|c|Y|}
\hline
\textbf{Noise sources} & \textbf{Time-domain characteristics} &
\textbf{Level} & \textbf{Spectrum characteristic} &
\textbf{Spectrum} & \textbf{Level} & \textbf{Solution} \\
\hline

RF-chain noise (Ref.) &
Uniform random noise (Peak: 0.015~V) &
Low &
Full-band random noise &
\adjustbox{max width=\linewidth, max height=2.2cm}{\includegraphics{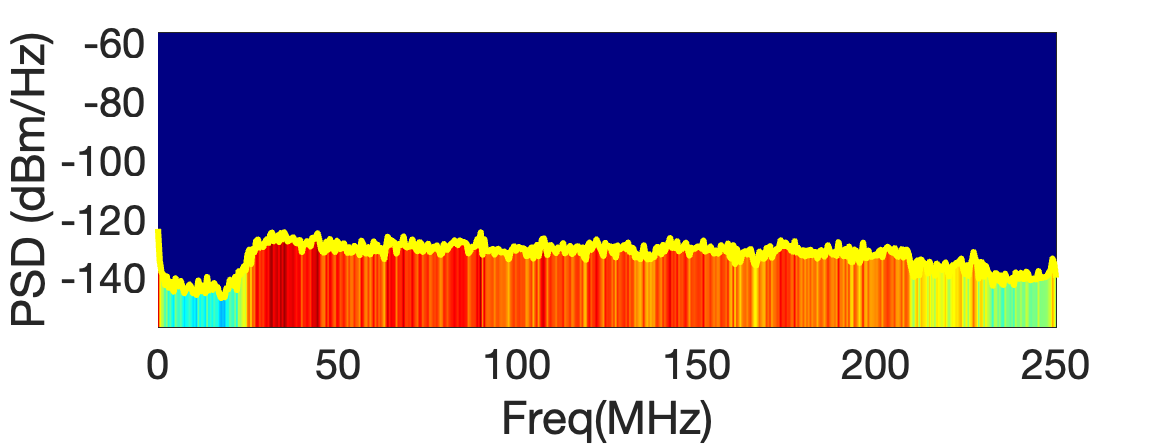}} &
Low & / \\
\hline

C\_PS (FEB power supply) &
Modulation small bump noise (Peak: 0.015~V) &
Low &
Bumps around 35, 110, and 190~MHz &
\adjustbox{max width=\linewidth, max height=2.2cm}{\includegraphics{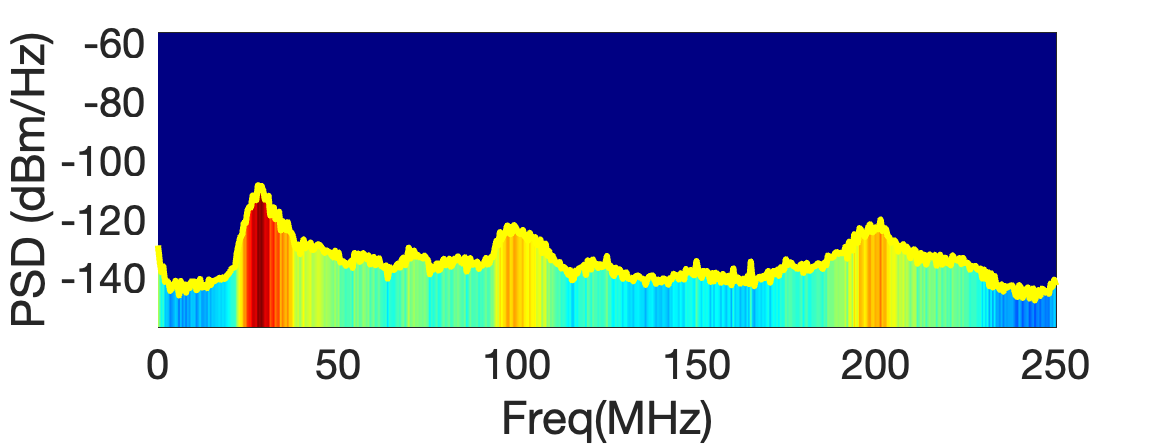}} &
Medium & Filter and twist shield cable \\
\hline

C\_CP/C\_BC (Charge controller) &
Pulses with high magnitude (Peak: 0.203~V) &
High &
Wide-band spectrum and low-frequency bump &
\adjustbox{max width=\linewidth, max height=2.2cm}{\includegraphics{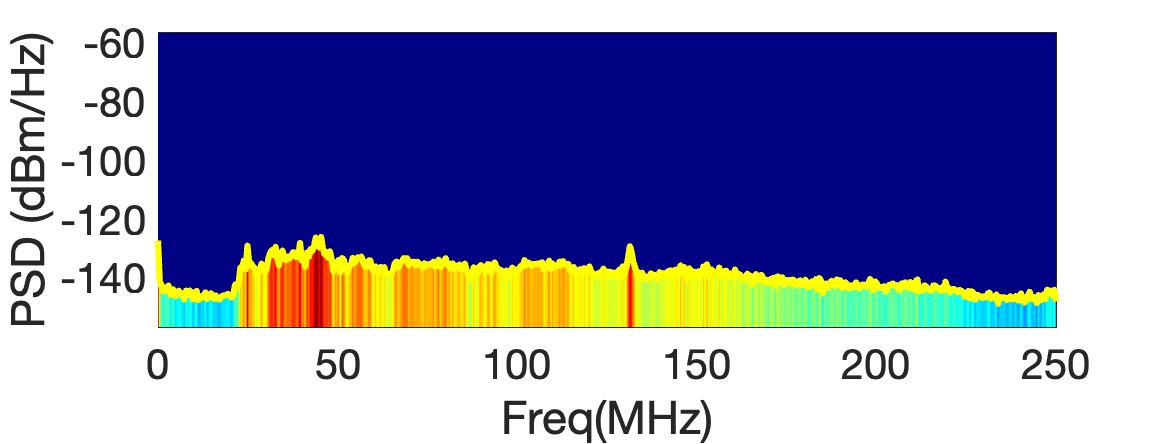}} &
Medium & Filter and twist shield cable \\
\hline

C\_WB (Wi-Fi Ethernet cable) &
Strong repeat small short pulses (Peak: 0.067~V) &
Medium &
Dense and continuous spectral contamination &
\adjustbox{max width=\linewidth, max height=2.2cm}{\includegraphics{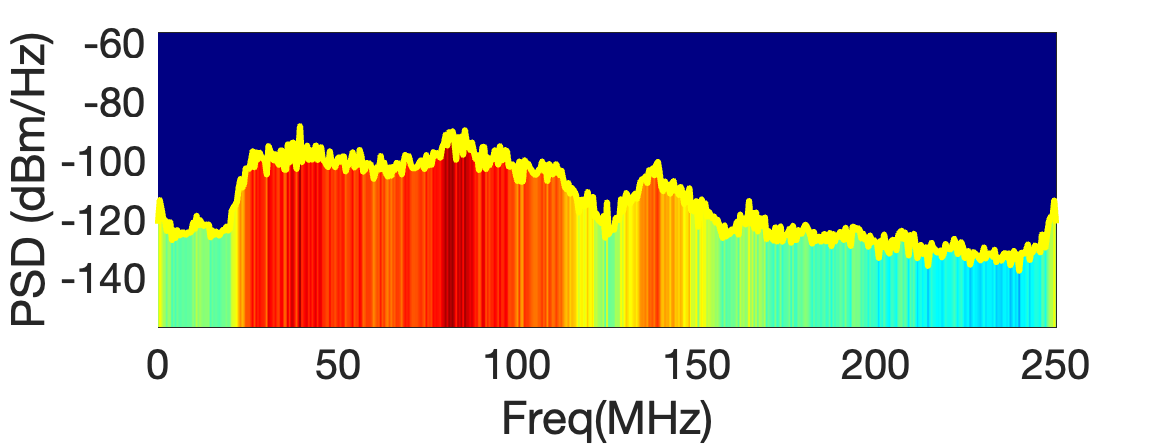}} &
High & Remove \\
\hline

C\_BA (Wi-Fi Bullet to antenna) &
Strong repeat small short pulses (Peak: 0.007~V) &
Medium &
Separated sharp spectral lines &
\adjustbox{max width=\linewidth, max height=2.2cm}{\includegraphics{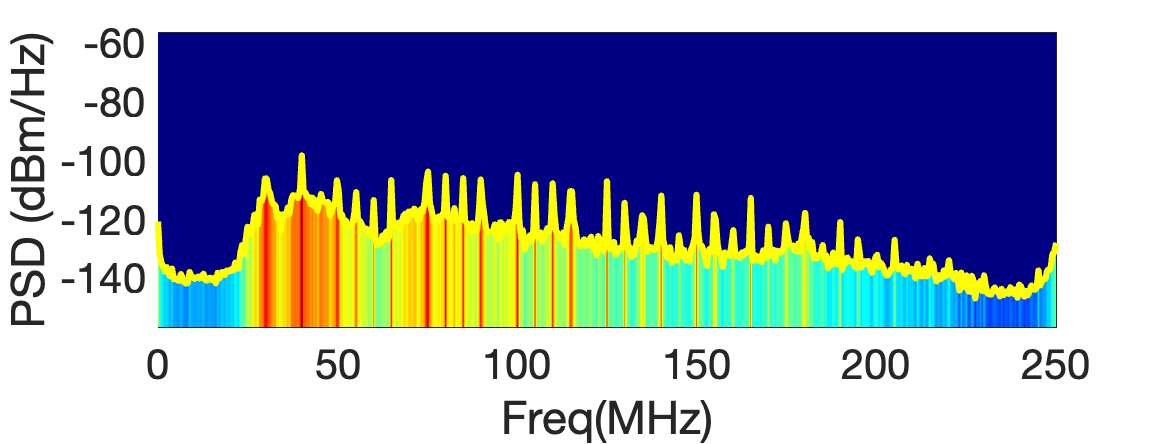}} &
Medium & Shield \\
\hline

C\_GPS (FEB to GPS antenna) &
Weak repeat short pulse (Peak: 0.009~V) &
Low &
Big bump around 25~MHz; small bump around 100/190~MHz &
\adjustbox{max width=\linewidth, max height=2.2cm}{\includegraphics{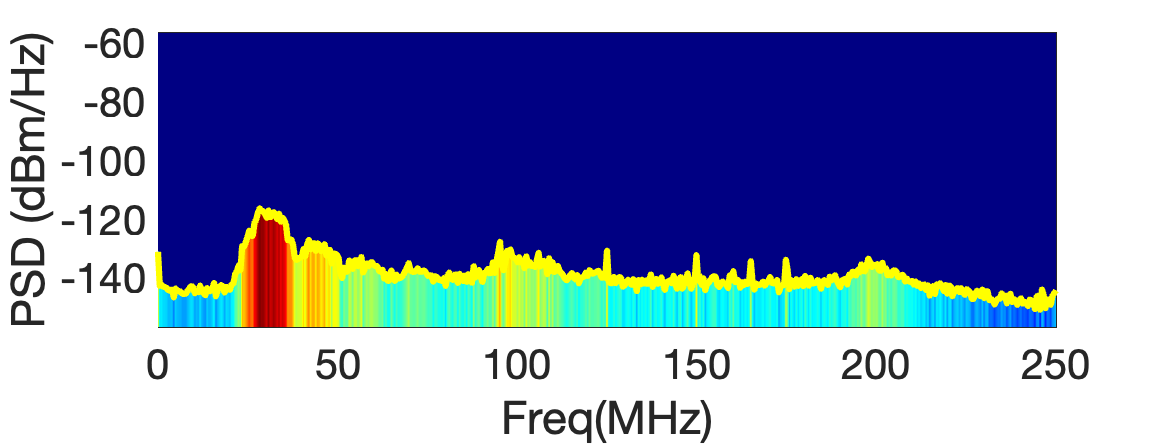}} &
Low & / \\
\hline

C\_Sensor (FEB to sensor) &
Weak repeat short pulse (Peak: 0.007~V) &
Low &
Dense and continuous contamination at 30--130~MHz &
\adjustbox{max width=\linewidth, max height=2.2cm}{\includegraphics{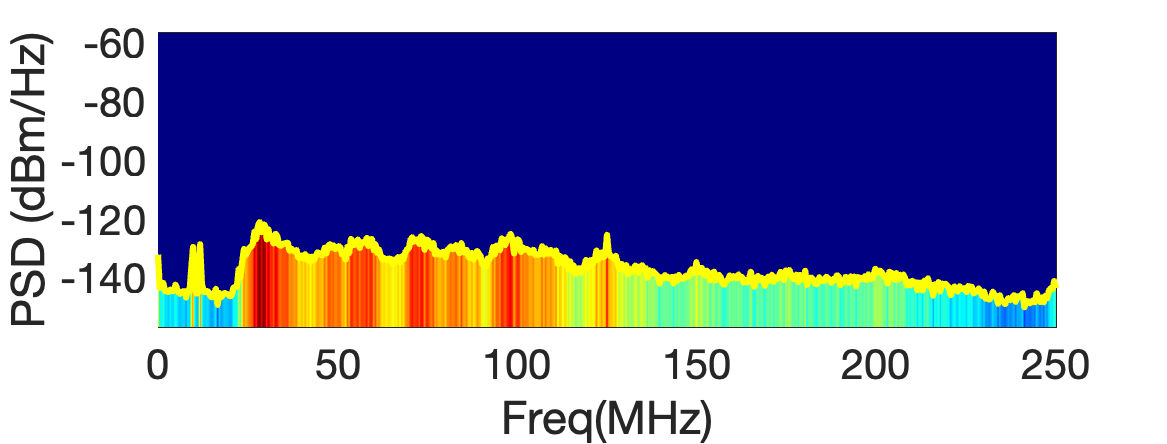}} &
Medium & Remove \\
\hline

\end{tabularx}
\end{table*}

\subsubsection{Cable Radiation Suppression}

The suppression of cable radiation can be achieved through the following strategies or their combinations: (i) embedding low-pass filters in through-wall connectors of cables and device boxes to block AC RFI currents flowing from devices to cables; (ii) installing magnetic ferrite rings for RF noise suppression; (iii) enclosing cables and components within metallic Faraday cages; (iv) using twisted-pair wiring for electromagnetic field cancellation; and (v) eliminating non-essential cabling. 

\begin{figure}[!t]
  \centering
  \subfloat[]{\includegraphics[width=0.45\linewidth]{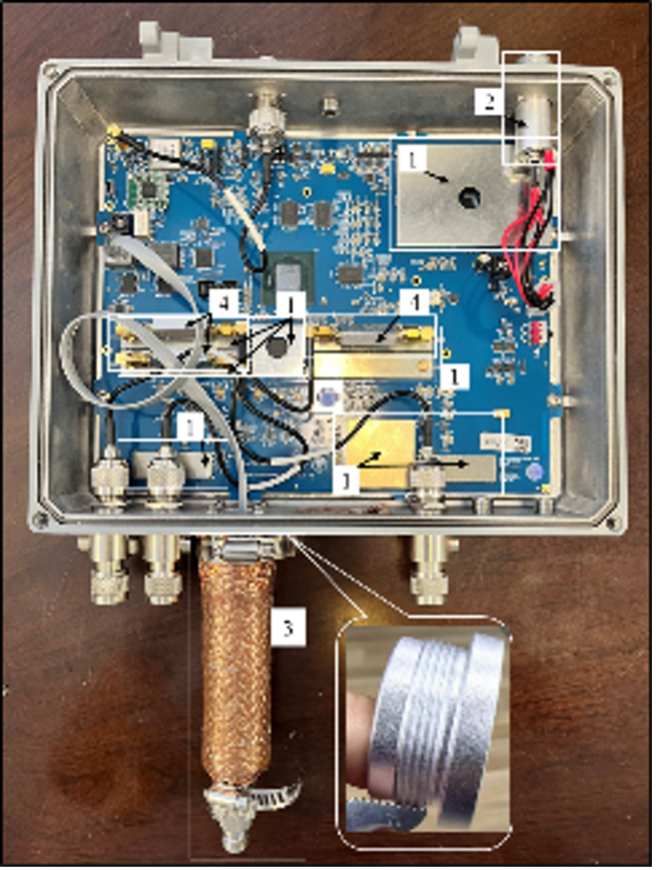}}
  \hfill
  \subfloat[]{\includegraphics[width=0.45\linewidth]{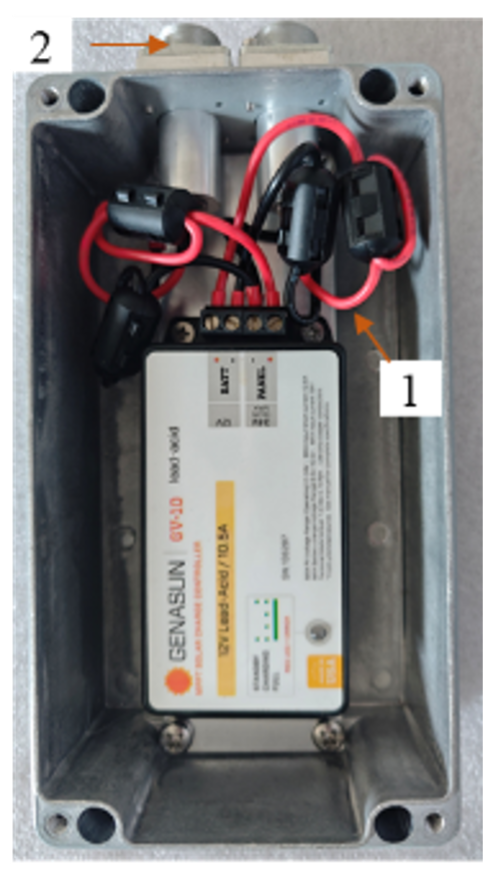}}\\[1ex]
  \subfloat[]{\includegraphics[width=0.45\linewidth]{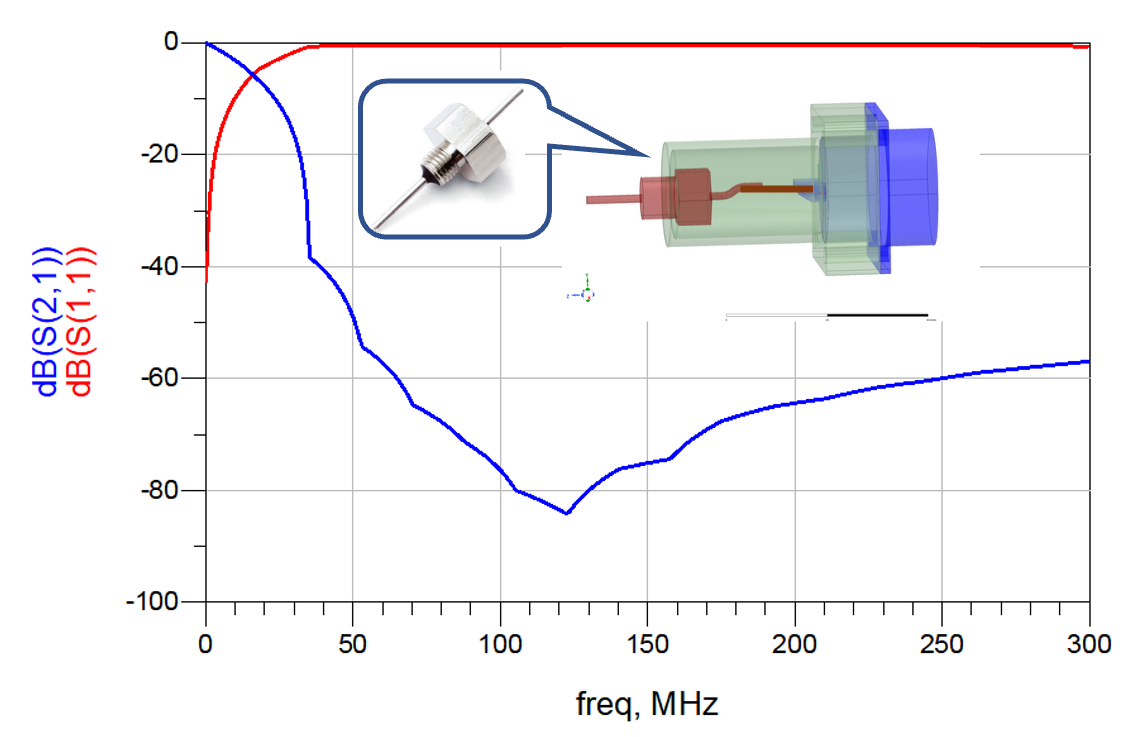}}
  \hfill
  \subfloat[]{\includegraphics[width=0.45\linewidth]{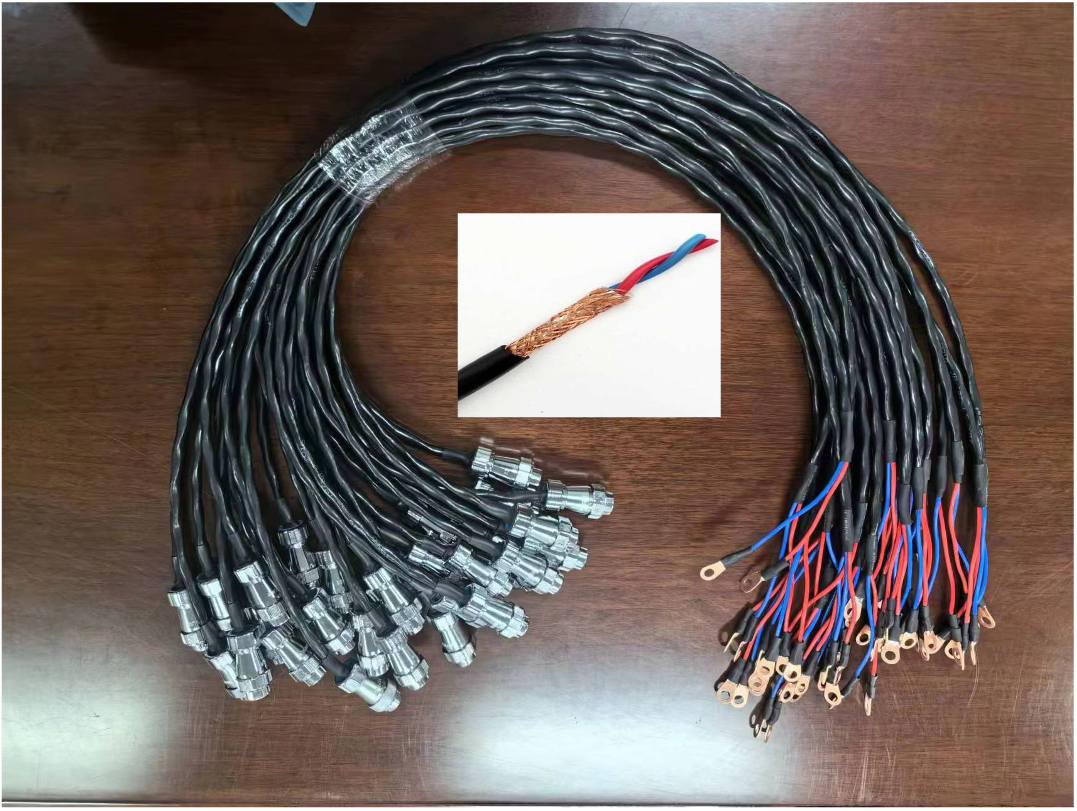}}
  \caption{Upgraded components for noise suppression: 
  (a) FEB box (metal cover of the Faraday cage removed for visibility); 
  (b) upgraded charge controller; 
  (c) performance of filters embedded in through-wall DC connectors; 
  (d) power cords using both copper mesh and twisted-pair methods.}
  \label{fig:upgrade}
\end{figure}

Mitigation measures for the high- and medium-risk items identified in Table~I include the following: 
-- \textbf{C\_WB (Wi-Fi Ethernet cable):} replacing the inbox cable + through-wall connector + outbox cable + Bullet with an inbox cable + metal adapter + Bullet, as shown in Fig.~\ref{fig:upgrade}(a), thereby eliminating the outbox Ethernet cable (corresponding to \#5 in Fig.~\ref{fig:test system}(b)).  
-- \textbf{C\_WA (Wi-Fi coaxial cable):} the stray noise contribution of the Wi-Fi cable was found to be strongly correlated with the current in the Wi-Fi transducer. To suppress this contribution, a copper-mesh Faraday cage was implemented, as shown by \#3 in Fig.~\ref{fig:upgrade}(a).

-- \textbf{C\_CP/C\_BC (charge-controller cables):} a three-tier suppression approach was implemented:  
(i) enclosing the controller in a metal box with through-wall DC connectors embedded with filters providing $>$45~dB suppression above 50~MHz (see Fig.~\ref{fig:upgrade}(c));  
(ii) replacing the original cables with shielded twisted-pair wiring to reduce EM radiation;  
(iii) installing ferrite rings (see \#1 in Fig.~\ref{fig:upgrade}(b)).  
The final charge-controller assembly is illustrated in Fig.~\ref{fig:upgrade}(b).  

-- \textbf{C\_PS (FEB power-supply cable):} strategies (i) and (ii) above were applied, as shown by \#2 in Fig.~\ref{fig:upgrade}(a).  

-- \textbf{C\_Sensor (sensor cable):} non-critical temperature and humidity sensors were removed from each detection unit.

The combined application of these mitigation strategies effectively suppresses cable-induced electromagnetic radiation, reducing non-RF-chain noise to levels below the sky-noise benchmark. This confirms that cable radiation constitutes a dominant but controllable noise source in self-triggered radio detector units.

\subsection{Suppression of Continuous Wave Interference from External Sources }

\begin{figure}[!t]
  \centering
  \includegraphics[width=3.4in]{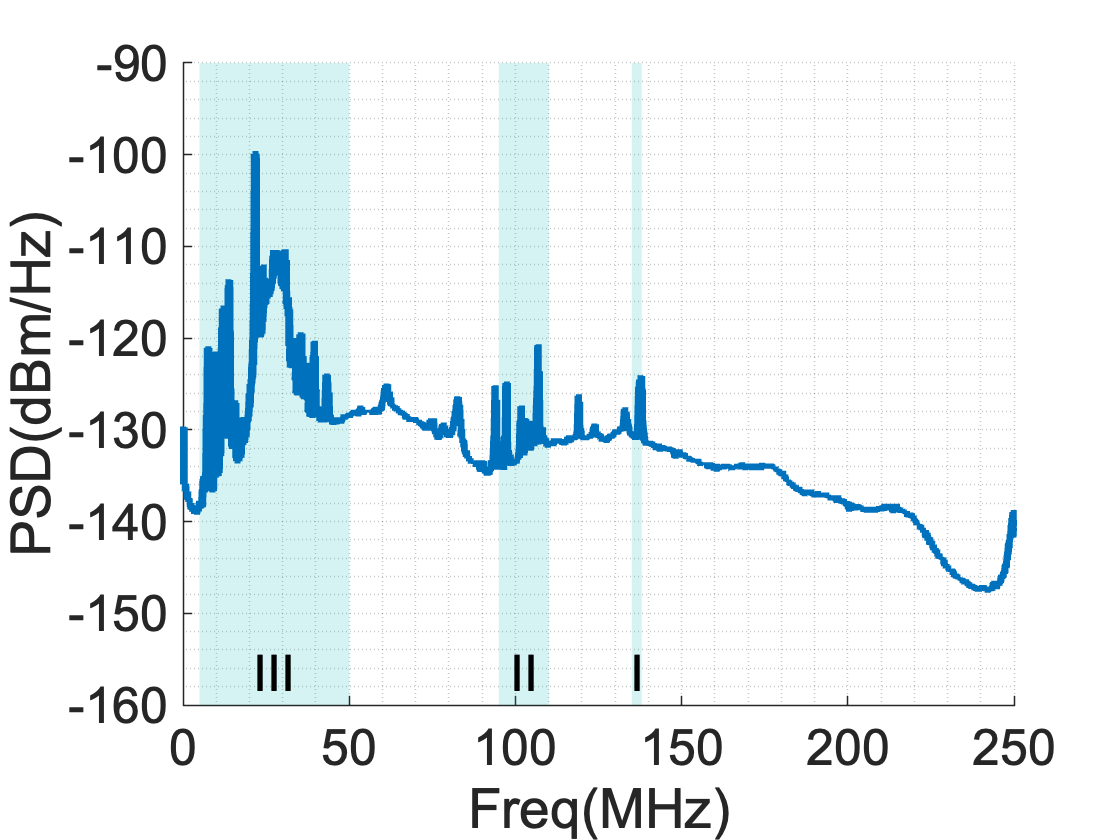}
  \caption{Typical AM, FM, and Satellite Signal Interference Spectrum}
  \label{fig:interference spectrum}
\end{figure}

After suppressing internal noise sources, residual interference observed during on-site measurements is dominated by external continuous-wave emissions. These signals primarily originate from broadcast and satellite services and require a combination of digital and analog mitigation strategies.

Fig.~\ref{fig:interference spectrum} reveals three dominant external interferences: (i) narrowband signals centered at 118 and 137~MHz; (ii) FM broadcast signals in the 88--108~MHz band; and (iii) AM-band emissions in the 5--45~MHz range. To mitigate interferences of types (i) and (ii), digital IIR filtering [Eq.~\eqref{eq:IIR}] was implemented in the FPGA of the FEB, leveraging its real-time processing capability~\cite{tan2012,Piskorowski2010,Lee2021}.

\begin{equation}
a_0 y[n] + a_1 y[n-1] + a_2 y[n-2] 
= b_0 x[n] + b_1 x[n-1] + b_2 x[n-2].
\label{eq:IIR}
\end{equation}

\begin{figure}[!t]
  \centering
  \subfloat[]{\includegraphics[width=3.4in]{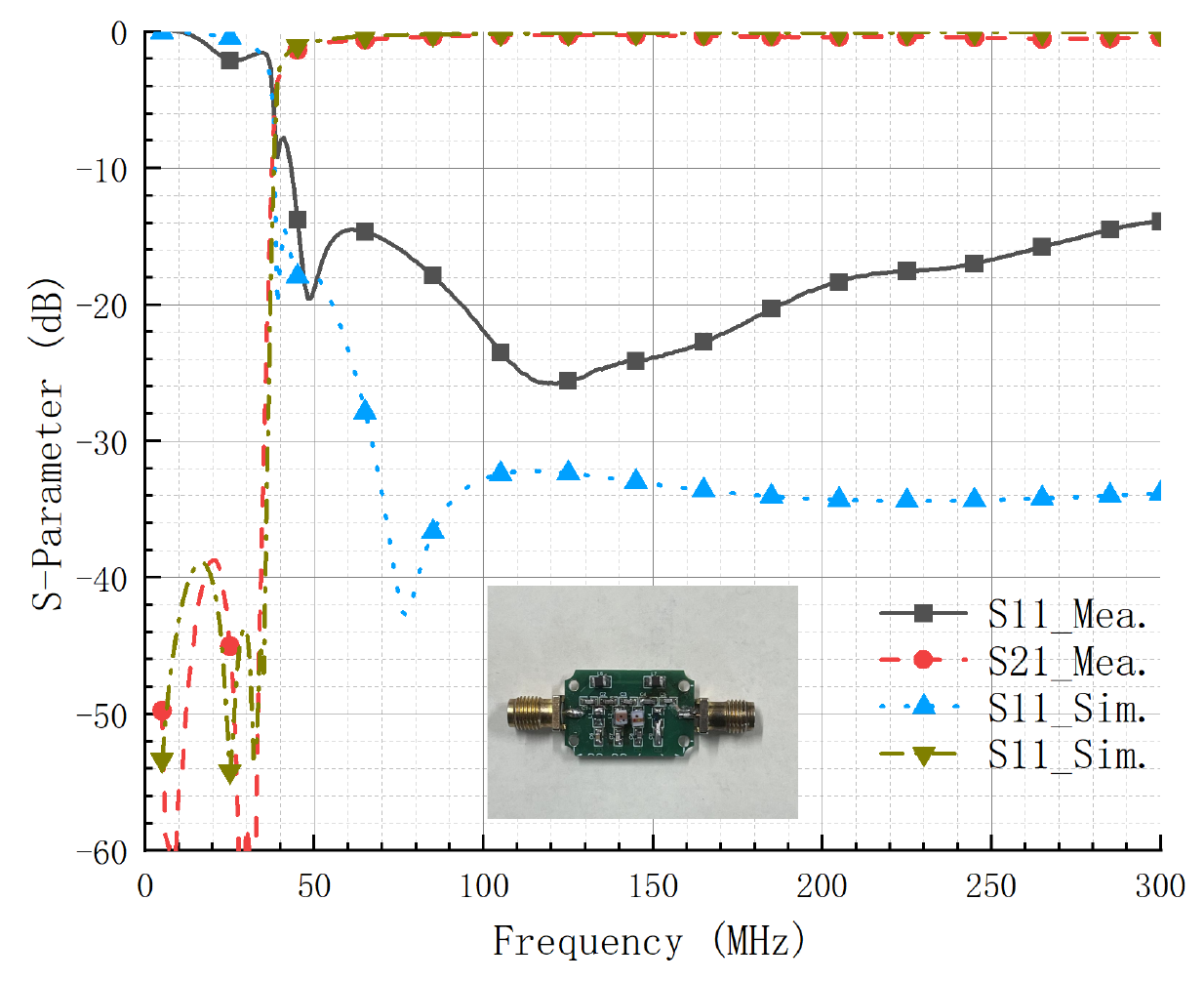}}
  \hfill
  \subfloat[]{\includegraphics[width=3.4in]{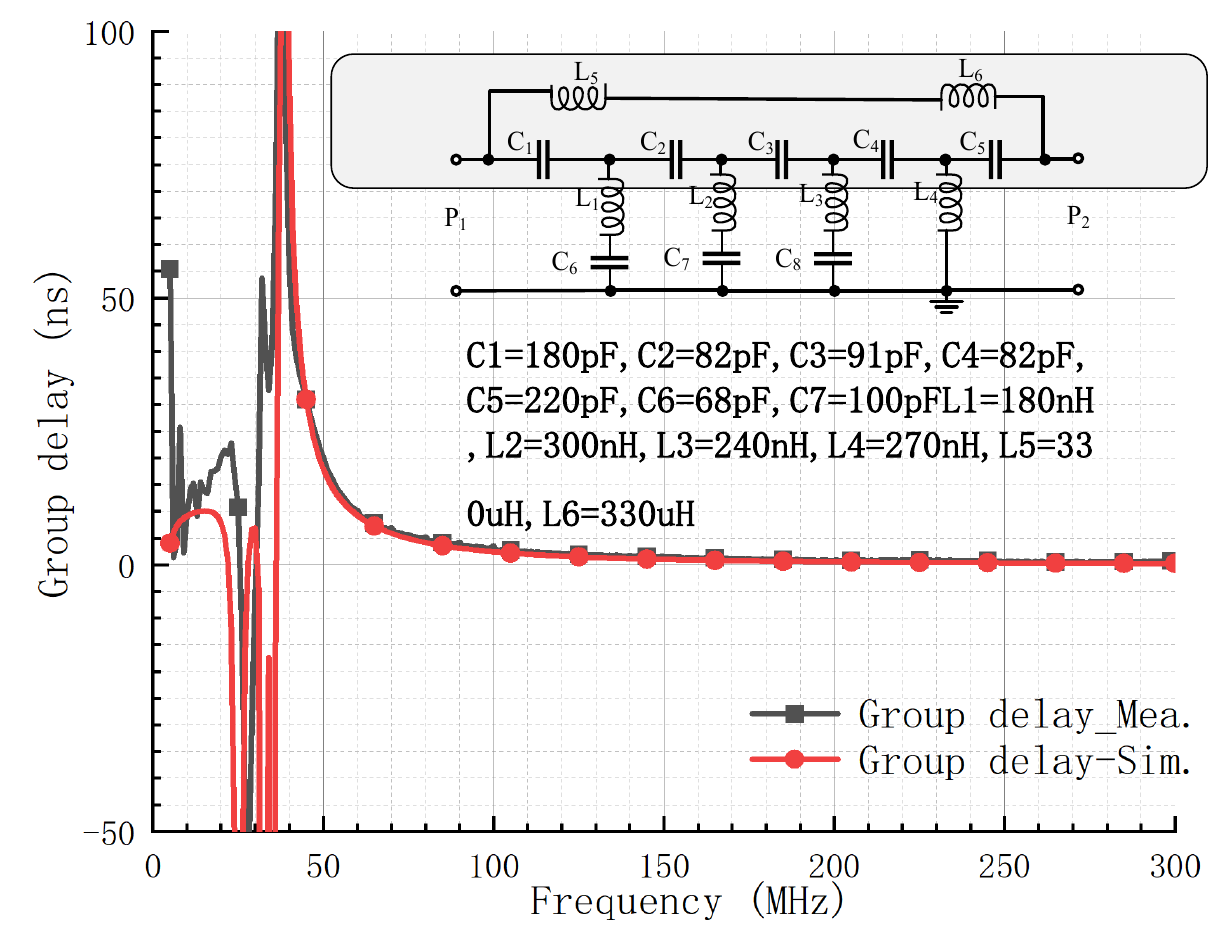}}\\[1ex]

\caption{High-impedance low-pass filter (HP-filter) for AM-band suppression: 
(a) photograph of the fabricated filter and measured $S$-parameters; 
(b) schematic diagram of the filter and corresponding group-delay response.}
\label{fig:HPfilter}

\end{figure}

For type-(iii) interference (AM band, 5--45~MHz), the combination of high amplitude and wide bandwidth necessitates a dedicated hardware solution. A high-impedance low-pass filter (HP-filter) was implemented between the LNA and VGA stages to suppress AM signals (see Fig.~\ref{fig:unit2}(a) and Fig.~\ref{fig:upgrade}(a)\#4). The design priorities were: (i) minimal in-band insertion loss; (ii) $>$40~dB out-of-band suppression; (iii) steep transition-band roll-off; (iv) flat group-delay response to preserve pulse shape~\cite{Scott2010}; and (v) an integrated DC-power path for LNA bias. Fig.~\ref{fig:HPfilter} shows the filter structure together with simulated and measured $S$-parameters. Validation confirms $>$40~dB AM suppression across 5--45~MHz with group-delay variation $<$10~ns within the passband, thereby maintaining pulse waveform integrity while effectively blocking interference.

At this stage, all identified noise sources have been analyzed and mitigated, except for transient-pulse interference, which will be addressed in Section~VI. Prior to that, the effectiveness of the proposed methods is verified through field tests presented in Section~V.

\section{Overall Background Noise Measurements}

A quantitative end-to-end validation of the detector sensitivity requires comparing the measured background noise with a baseline prediction derived from the sky-noise model (Sec.~III) and the RF-chain noise budget (Sec.~IV). In Fig.~\ref{fig:system_noise}(a), the purple curve denotes the predicted internal system noise, while the yellow dashed curve represents the sky background contribution. Power summation yields the red curve, corresponding to the expected total noise level at the ADC input. Although accessory-induced noise was substantially suppressed (cf. Fig.~\ref{fig:noise EMC}), residual antenna-coupled contributions cannot be entirely eliminated and may slightly elevate the measured noise floor above the ideal prediction. Quantifying this residual component independently is difficult because dedicated anechoic characterization around 50~MHz is not available; therefore, the validated total background level is expected to lie marginally above the red curve.

Following standard observational protocols, the test system recorded background data continuously over a 24-h period under stable operating conditions. Data segments contaminated by satellites, FM broadcasts, or other identifiable interferences were excluded to construct RFI-minimized background datasets. Spectral averaging of these curated data yields the measured background noise spectrum shown as the blue curve in Fig.~\ref{fig:system_noise}(a). The predicted combined-noise model (red curve) agrees with the measurement within approximately 1~dB, validating both the sky-noise-based prediction methodology and the implemented noise-suppression measures. Beyond spectral agreement, the achieved noise suppression can be quantified directly in the time domain.
The root-mean-square (RMS) value of the background waveform, computed from the ADC time-series data under identical analysis conditions, is reduced from an average of 70.58~ADC counts in an early prototype configuration to 11.66~ADC counts in the optimized system presented here.
This corresponds to a reduction by a factor of approximately six in RMS amplitude, or more than 15~dB in noise power, providing a clear and quantitative measure of the system-level improvement achieved through the integrated RF-chain optimization and EMC mitigation strategy.Adopting a standard self-trigger condition of five times the RMS noise level, applied independently to each polarization channel, the corresponding trigger threshold can be expressed in terms of an equivalent incident electric-field strength.
Using the measured antenna response together with the calibrated RF-chain transfer function, the optimized system reaches an effective trigger threshold of approximately 75~$\mu$V\,m$^{-1}$.
For comparison, the early prototype configuration corresponds to a threshold of about 450~$\mu$V\,m$^{-1}$ under the same criterion.
This reduction by nearly one order of magnitude at the field level directly reflects the improved background-limited sensitivity of the detector unit.Importantly, within the core sensing band for EAS self-triggering (50--120~MHz), the measured noise floor remains below the Galactic background level, demonstrating background-limited sensitivity in the intended operational band.

Fig.~\ref{fig:system_noise}(b) and (c) further compare simulated and measured 24-h time evolution of the integrated power in the 60--90~MHz band for the EW and NS channels, respectively. The simulated reference includes the sky background and the internal system noise (red solid curves), while the measurements are shown as blue solid curves. The observed diurnal modulation follows the expected variation induced by Earth's rotation relative to the Galactic plane. A residual deviation of approximately 0.7~dB between prediction and measurement is observed, which is consistent with incomplete suppression of residual antenna-coupled noise contributions discussed above.

Together with the RF-chain response calculations in Sec.~III-B, these measurements establish and validate a consistent simulation-to-data pipeline for the GRAND detector units. This pipeline links sky-noise modeling and RF-chain transfer characterization to the ADC-level background observed in the field, providing a verified quantitative baseline for subsequent self-trigger and interference-rejection studies.

The reduction of the effective electric-field trigger threshold has direct consequences for the physics reach of a self-triggered array.
Assuming a standard coincidence requirement in which signals from at least five antenna units are required to define a detected event, the accessible cosmic-ray energy range is extended from approximately $10^{17.6}$--$10^{18}$~eV for the early prototype configuration to about $10^{16.8}$--$10^{18}$~eV for the optimized system presented here.
Based on an event-rate estimate using the TALE cosmic-ray flux model, this improvement translates into an increase of the expected daily detection rate from fewer than ten events per day to roughly two hundred events per day.
While the detailed rate-estimation methodology will be presented in a dedicated publication, these figures provide an order-of-magnitude quantification of the scientific impact enabled by achieving a Galactic-noise-limited detector performance.
This dramatic gain directly reflects the steeply falling cosmic-ray energy spectrum, for which lowering the trigger threshold yields a strongly non-linear increase in event statistics.

\begin{figure}[!t]
  \centering
  \includegraphics[width=3.4in]{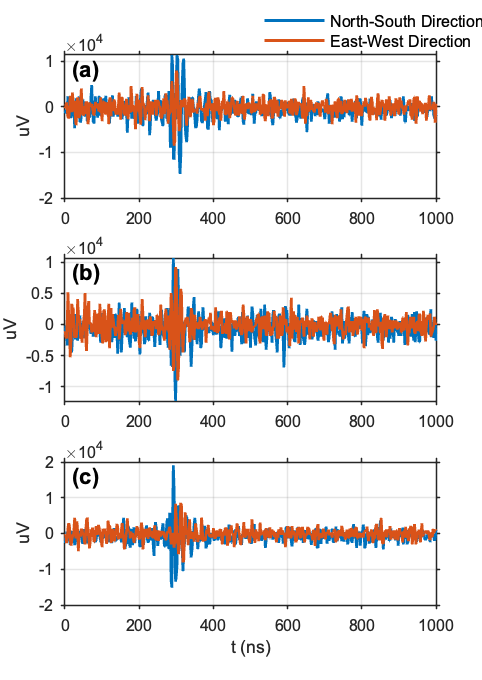}
  \caption{Simulated ADC-level time-domain radio waveforms on the East--West and North--South antenna arms for three cosmic-ray-induced EAS events. The primary energies are (a) 0.699 EeV, (b) 0.423 EeV, and (c) 0.859 EeV. The corresponding arrival directions are $\theta = 63.2^\circ$, $\phi = 286.6^\circ$ for (a); $\theta = 53.1^\circ$, $\phi = 69.6^\circ$ for (b); and $\theta = 60.2^\circ$, $\phi = 113.9^\circ$ for (c).}
  \label{fig:rfchain result}
\end{figure}

\begin{figure}[!t]
  \centering
  \includegraphics[width=3.4in]{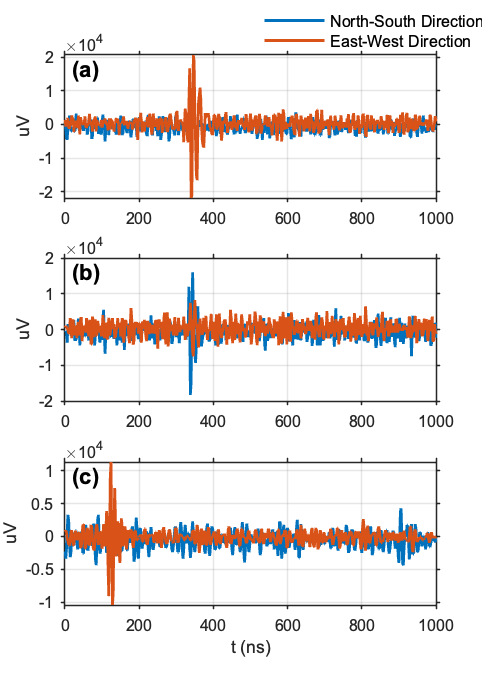}
  \caption{Examples of measured time-domain radio waveforms recorded by GRAND antennas for three cosmic-ray candidate events. The reconstructed arrival direction for (a) is $\theta = 80.5^\circ$ and $\phi = 185.9^\circ$. For (b), the reconstructed arrival direction is $\theta = 76.3^\circ$ and $\phi = 82.7^\circ$. For (c), the reconstructed arrival direction is $\theta = 76.2^\circ$ and $\phi = 356.6^\circ$.}

  \label{fig:measured_waveform}
\end{figure}

As an illustration of expected self-trigger conditions, Fig.~\ref{fig:rfchain result} shows simulated ADC-level time-domain radio waveforms on the East--West and North--South antenna arms for three cosmic-ray-induced EAS events with different energies and arrival directions. These examples illustrate the diversity of waveform morphologies under different shower geometries and energies, as well as the event-dependent signal amplitudes and pulse shapes in the EW and NS channels due to geomagnetic polarization. We emphasize that Fig.~\ref{fig:rfchain result} provides simulation-based illustrations rather than confirmed measured events. Three examples of measured triggered waveforms recorded by the GRAND system are shown in Fig.~\ref{fig:measured_waveform}.

Galactic noise represents an irreducible stationary background for omnidirectional low-frequency radio detectors. Once the internal electronic noise is suppressed below this level, the achievable sensitivity is fundamentally limited by the sky background itself. 
Given the typical radio-emission intensity of extensive air showers and the simulated ADC-level pulses shown in Fig.~\ref{fig:rfchain result}, this background level translates into a practical detection threshold around $10^{16.5}$~eV. Consequently, the present detector and radio detection techniques in general are intrinsically better suited for the study of the highest-energy cosmic particles.
Direct quantitative comparison with other EAS radio experiments is not straightforward, as differences in antenna directivity, field of view, trigger logic, and RF-chain architecture significantly affect the effective noise environment. While the periodic Galactic emission is commonly used for amplitude calibration in experiments such as AERA~\cite{AERAref} and TREND~\cite{charrier2019}, none of them has presented an equally detailed, system-level decomposition and measurement-driven validation of individual noise contributions throughout the complete self-triggered RF chain.

\section{Brief Introduction to the Rejection of Pulse Interference}

As illustrated in Fig.~\ref{fig:Schematic diagram}, the self-triggered detection system deployed at the GRANDProto300 site is exposed to multiple classes of transient pulse interference.

More specifically, the discrimination between anthropogenic interference and EAS radio signals is achieved at the system-response level rather than relying on a single algorithmic criterion.
Man-made radio-frequency interference typically manifests as either continuous-wave narrowband emissions or repetitive transient pulses with characteristic spectral features.
In contrast, EAS-induced radio signals exhibit broadband spectra, nanosecond-scale impulsive waveforms, and polarization patterns governed by geomagnetic and Askaryan emission mechanisms~\cite{chiche2022}.
In this work, we design the RF chain to operate in a regime limited by the irreducible Galactic sky noise.
Continuous-wave interference is suppressed through a combination of analog filtering and FPGA-level digital processing.
Together with measurement-driven characterization of waveform-level temporal and spectral features, this approach provides a physical system-level basis for reducing anthropogenic interference while preserving impulsive EAS radio signals.
In this paper, we focus on the detector-unit design, quantitative noise reference, and measurement-validated suppression of dominant internal and continuous-wave interference; detailed trigger-level pulse-classification algorithms are beyond the scope of this work and will be reported separately.

During test operations, such pulses were observed to trigger the acquisition system at rates reaching $\sim$1~kHz, far exceeding the expected rate of ultra-high-energy cosmic-ray events.
Detailed investigations identified the dominant contributors as electrostatic discharges associated with civil aircraft, electromagnetic interference (EMI) emitted by a power-transmission facility located approximately 15~km from the site, and, to a lesser extent, solar energetic particle (SEP) activity.

Time- and frequency-domain analyses reveal that most interference pulses exhibit characteristic features that differ from those expected for EAS-induced radio signals. For instance, EMI originating from the power-transmission facility shows enhanced spectral power above approximately 150~MHz, whereas aircraft-related discharges display distinct spectral structures, potentially influenced by airframe geometry and discharge dynamics. These differences provide a physical basis for early-stage discrimination between anthropogenic interference and genuine EAS candidates.

For interference dominated by high-frequency components, a computationally efficient online mitigation strategy consists of applying a digital finite impulse response (FIR) high-pass filter~\cite{thamizharasan2023}. While this approach inevitably reduces sensitivity to a subset of broadband EAS signals, it provides robust suppression of dominant interference sources at the trigger level and thus significantly reduces the data acquisition load.

\begin{equation}
H(z) = \sum_{k=0}^{N-1} b_k z^{-k}
\;\;\Rightarrow\;\;
y[n] = \sum_{k=0}^{N-1} b_k \, x[n-k].
\label{eq:FIR}
\end{equation}

\begin{table}[!t]
\caption{150~MHz High-pass FIR Filter Coefficients ($N=74$)}
\label{tab:FIR}
\centering
\renewcommand{\arraystretch}{1.2}
\begin{tabular}{|c|c|c|c|c|c|}
\hline
$b_0$ & $b_1$ & $b_2$ & $b_3$ & $b_4$ & $b_5$ \\
\hline
-0.0021 & 0.0104 & 0.0157 & 0.0086 & -0.0041 & -0.0055 \\
\hline
$b_6$ & $b_7$ & $b_8$ & $b_9$ & $b_{10}$ & $\cdots$ \\
\hline
0.0038 & 0.0058 & -0.0035 & -0.0068 & 0.0030 & $\cdots$ \\
\hline
\end{tabular}
\end{table}

Beyond FIR-based suppression, several complementary strategies can be employed to further enhance pulse-interference rejection. These include: 
(i) \textit{statistical discrimination enhancement}, in which accumulated interference statistics are used to refine trigger thresholds and classification criteria; 
(ii) \textit{template fitting}, where measured transient waveforms are compared to simulated EAS templates, an approach already demonstrated in offline analyses within the GRAND framework~\cite{correa2025}; and 
(iii) \textit{adaptive denoising}, involving the development of targeted algorithms that integrate time--frequency characteristics with offline event reconstruction.

In addition, emerging machine-learning techniques, in particular convolutional neural network (CNN) classifiers exploiting multi-dimensional waveform features, have demonstrated strong potential for pulse identification and classification. Such approaches are currently under active investigation within the GRAND collaboration~\cite{hare2023,benoit2025}. A dedicated companion publication will present a comprehensive description and validation of these methods.

\section{Conclusion}

Noise and transient interference constitute fundamental challenges for self-triggered low-frequency radio detection systems operating in the classical 30--350~MHz band. In this work, we have presented a systematic, end-to-end analysis of noise and interference sources, their coupling paths, and injection mechanisms within a representative self-triggered array architecture. Based on a quantitative sky-background response model at the single-unit level, explicit noise-suppression targets were defined, providing a physically motivated benchmark for system design.

Guided by these benchmarks, we developed and validated a co-designed hardware--software mitigation framework encompassing RF-chain optimization, electromagnetic compatibility (EMC) treatments, suppression of self-generated non-RF noise, and dedicated continuous-wave interference filtering. Field measurements demonstrate good spectral and temporal agreement with simulations, and confirm that the detector units operate in a background-limited regime where the residual system noise approaches the Galactic sky-noise level within the core observational band.

Quantitatively, the optimized detector unit reduces the time-domain RMS background noise from 70.58 to 11.66~ADC counts compared to an early prototype, corresponding to a reduction of more than 15~dB in noise power. This improvement lowers the effective self-trigger threshold from approximately 450 to 75~$\mu$V\,m$^{-1}$ and extends the expected self-triggered detection threshold down to cosmic-ray energies of order $10^{16.8}$~eV, while operating in a Galactic-noise-dominated regime within the 50--120~MHz band.

In addition, simulation-based studies incorporating measured noise spectra show that the developed system satisfies the signal-to-noise requirements for self-triggered detection of extensive air showers in the targeted energy range. Site-specific pulse-interference characterization and mitigation strategies validated at the GRANDProto300 site further demonstrate the robustness of the proposed approach under realistic operating conditions.

Overall, this work establishes a coherent system-level methodology for designing, characterizing, and validating self-triggered radio detector units under realistic noise and interference conditions. The presented techniques and design principles provide a solid technical foundation for future large-scale radio arrays for ultra-high-energy cosmic-ray and neutrino detection, and are readily transferable to other low-frequency sensing applications facing similar electromagnetic challenges.

\section*{Acknowledgment}
The authors are grateful to the GRAND Collaboration for support, and in particular to the Radboud group led by Charles Timmermans for designing the GRAND FEB board, which served as an essential test platform for this study. 
The views expressed in this paper are solely those of the authors and do not represent those of the GRAND Collaboration.

\end{document}